\documentclass[aps,superscriptaddress,twocolumn,twoside,floatfix,pra,nofootinbib,a4paper]{revtex4-2}

\usepackage[charter,cal=cmcal,sfscaled=false]{mathdesign}

\usepackage{dcolumn}
\usepackage{bm}
\usepackage{dsfont}
\usepackage{graphicx,epsfig}
\usepackage{amsmath}
\usepackage{braket}
\usepackage{caption}
\usepackage{float}
\usepackage[shortlabels]{enumitem}
\usepackage{mathtools}
\usepackage{physics}

\usepackage{blindtext}
\usepackage{color}

\usepackage[colorlinks]{hyperref}
\hypersetup{
	colorlinks = true,
	urlcolor = {blue},
	citecolor = {blue},
	linkcolor= {blue}
}

\usepackage[bottom]{footmisc}

\renewcommand{\eqref}[1]{Eq.~(\ref{#1})}
\newcommand{\figref}[1]{Fig.~\ref{#1}}
\newcommand{\appref}[1]{App.~\ref{#1}}
\newcommand{\secref}[1]{Sec.~\ref{#1}}

\begin{document}

\title{Sequential Semi-Device-Independent Quantum Randomness Certification}

\author{Carles Roch i Carceller}\email{carles.roch\_i\_carceller@teorfys.lu.se}\address{Physics Department and NanoLund, Lund University, Box 118, 22100 Lund, Sweden}

\author{Hanwool Lee}\address{Faculty of Information Technology, University of Jyväskylä, 40100 Jyväskylä, Finland}

\author{Jonatan Bohr Brask}\address{Center for Macroscopic Quantum States bigQ, Department of Physics, Technical University of Denmark, Fysikvej 307, 2800 Kgs.~Lyngby, Denmark}

\author{Kieran Flatt}\address{School of Electrical Engineering, Korea Advanced Institute of Science and Technology (KAIST), 291 Daehak-ro, Yuseong-gu, Daejeon 34141, Republic of Korea}

\author{Joonwoo Bae}\address{School of Electrical Engineering, Korea Advanced Institute of Science and Technology (KAIST), 291 Daehak-ro, Yuseong-gu, Daejeon 34141, Republic of Korea}

\begin{abstract}
Quantum measurements under realistic conditions reveal only partial information about a system. Yet, by performing sequential measurements on the same system, additional information can be accessed. We investigate this problem in the context of semi-device-independent randomness certification using sequential maximum confidence measurements. We develop a general framework and versatile numerical methods to bound the amount of certifiable randomness in such scenarios. We further introduce a technique to compute min-tradeoff functions via semidefinite programming duality, thus making the framework suitable for bounding the certifiable randomness against adaptive attacking strategies through entropy accumulation. Our results establish sufficient criteria showing that maximum confidence measurements enable the distribution and certification of randomness across a sequential measurement chain.
\end{abstract}

\maketitle

\section{Introduction}

The inherent unpredictability of quantum phenomena makes quantum systems a natural source of randomness. This fundamental feature has driven key advances in quantum information, particularly in cryptography, where certified randomness offers the strongest security guarantees \cite{gisin2002,pirandola2020,mannalatha2023}. Developing practical and secure protocols for quantum randomness certification remains a central goal in the field. Achieving this balance, however, is not trivial. On one hand, Bell nonlocality in multipartite scenarios reveals probability distributions that quantum systems can generate but classical systems cannot \cite{PhysicsPhysiqueFizika.1.195, RevModPhys.86.419}. This not only provides a fundamental tool to rule out local hidden variable models but also serves as a key resource for practical applications, such as device-independent quantum cryptography \cite{barrett2005,PhysRevLett.98.230501,vazirani2014} and randomness generation \cite{pironio2010,miller2016,acin2016}. On the other hand, semi-device-independent prepare-and-measure scenarios offer practical frameworks, lowering experimental demands of Bell violations by adding few assumptions on the internal workings of the devices, while still harnessing quantum unpredictability \cite{pawlowski2011,li2012,zhou2016,pivoluska2021,mannalath2022,carceller2022}.

Recently, both Bell nonlocality and state discrimination have been considered in a sequential scenario, where a receiver applies non-destructive measurements and sends the resulting state to the next receiver, and so on \cite{Gallego_2014}. Sequential quantum scenarios are generally nontrivial since quantum properties, such as the monogamy of nonlocal correlations and the impossibility of copying quantum states, limit an extension to a multi-party network scenario \cite{PhysRevA.61.052306}. It is worth mentioning that monogamy and the no-cloning theorem, as well as the impossibility of perfect state discrimination, are in fact interrelated as consequences of the fact that quantum theory is non-signalling and contains Bell violations \cite{PhysRevA.73.012112}. In sequential Bell violations, a single party shares nonlocal correlations with arbitrarily many others \cite{PhysRevLett.114.250401, PhysRevLett.125.090401}. Sequential state discrimination has also been studied in terms of unambiguous discrimination \cite{bergou2013} as well as its generalization, maximum-confidence discrimination \cite{hanwool2024}, which also includes minimum-error discrimination as a particular case. Note that in these cases, Bell violations and the rate of conclusive outcomes for state discrimination are just above the thresholds, given which, i.e., weak Bell violations and high rates of inconclusive outcomes, the certification of randomness of each party in the sequential scenario is highly non-trivial and little is known yet along this line.

In this work, we establish a framework for certifying randomness in a sequential state discrimination setting, where assumptions are less stringent than those required for Bell violations. We consider realistic scenarios where measurements may not function as desired---yielding undetected or inconclusive events---and thus cannot be trusted. We present theoretical tools to certify randomness in a realistic sequential scenario. Then, we take two approaches. One is a measurement-device-independent scenario, where all states---including post-measurement states---are trusted, but measurements are completely uncharacterized. The second approach relaxes the previous by removing all trust from post-measurement states and only maintaining minimal assumptions on the initial state preparations, such that these are initially pure and a bound on their overlap is provided. We show that sequential randomness can be certified in both cases. These results can be rephrased as the distribution of randomness inherent in a measurement of an ensemble among multiple parties. Randomness may originate from the fact that no measurement can perfectly discriminate non-orthogonal states, hence, identified as the indistinguishability of quantum states. The randomness is given to a single party by the sender, and the receiver can distribute the randomness by applying non-destructive measurements and sending remaining states to the next party, and so on. Our results show that randomness distributed in a sequential scenario can be certified.

The remainder of this work is organized as follows. In \secref{sec:seq_PAM} we present the sequential prepare-and-measure scenario with two consecutive measurements that we investigate, and the correlations that arise in. We continue in \secref{sec:seq_qrng} introducing the randomness certification framework. We provide the semidefinite programming tools to compute the bounds on the min- and Shannon entropies in both approaches, i.e.~with fully or partially trusted preparations. We deploy the methods to compute bounds on the certifiable randomness in \secref{sec:res} considering the simplest case of two state preparations and two sequential measurements performing maximum confidence measurements. We observe that in some cases, it is possible to simultaneously distribute the certifiable randomness in both sequential measurements, which leads us to finding the necessary condition to be able to certify randomness in $N$ sequential measurements in \secref{sec:n_seqs}. We end in \secref{sec:disc} with an outlook of our methods and discussing the importance of our results.

\section{Sequential prepare-and-measure} \label{sec:seq_PAM}

\begin{figure}
	\centering
    \includegraphics[width=0.45\textwidth]{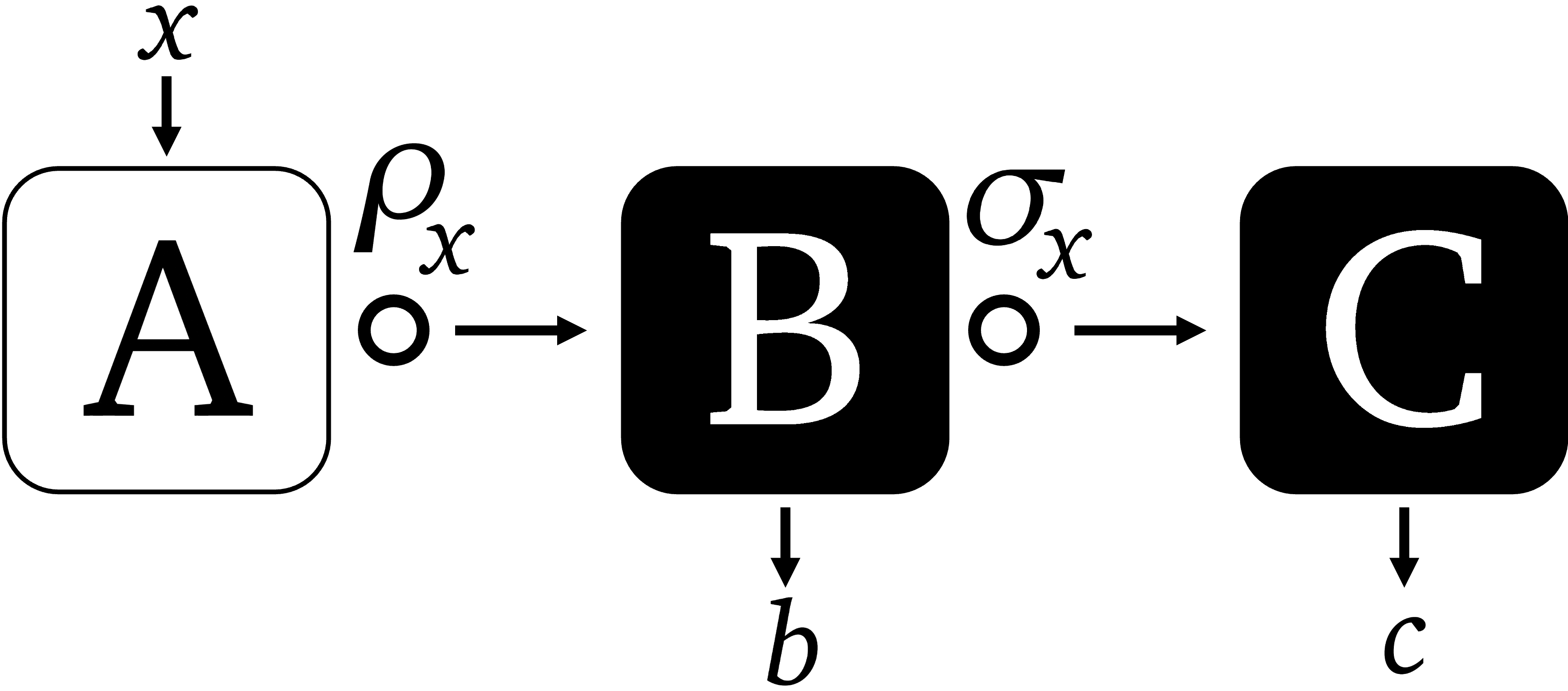}
  \caption{\textit{Sequential prepare-and-measure framework.} Alice (A) encodes a symbol $x$ into a quantum state $\rho_x$. This is sent to Bob (B), observing a measurement outcome $b$. Lastly, Charlie (C) measures Bob's post-measurement state $\sigma_x$ and produces an outcome $c$.}
 \label{fig:scenario}
\end{figure}

Consider the sequential prepare-and-measure scenario depicted in \figref{fig:scenario}. With probability $p_x$, Alice encodes a classical symbol $x\in X$ into a quantum state $\rho_x$ which is sent to Bob, who performs a measurement $\{M_{b}\}$ and receives $b$ as a measurement outcome. After doing so, Bob will possess the remaining post-measurement state 
\begin{align} \label{eq:pm_states}
\sigma_x := \sum_b K_b^{\dagger} \rho_x K_b \ ,
\end{align}
where $K_b$ are the Kraus operators corresponding to Bob's instrument, i.e~$M_{b}=K_b K_b^{\dagger}$. In general, Bob’s measurement may not extract all the information encoded in Alice’s states. The remaining quantum state $\sigma_x$ can thus still contain information about $x$ that can be accessed by a subsequent observer, Charlie, who performs a measurement $\{N_c\}$ with outcomes $c$. This leads to a natural trade-off: the more information Bob gains about $x$, the less remains available to Charlie, and vice versa.

We study frameworks with completely untrusted measurements. Specifically, we assume that all measurement devices have been pre-programmed to perform certain unknown measurement strategies labeled by $\lambda$ according to a distribution $q(\lambda)$. The preparation is partially characterised. We assume that the prepared states are pure with a fixed overlap in every round, $\left|\braket{\psi_x}{\psi_{x'}}\right| = \delta$. Note that, due to unitary invariance, without loss of generality we can take the states to be independent of $\lambda$. 
Therefore, the correlations that can be observed in this scenario can be written as
\begin{align}
p(b,c|x) = \sum_\lambda q(\lambda) \Tr\left[\rho_x K_b^{(\lambda)}N_{c}^{(\lambda)}K_b^{(\lambda)\dagger}\right] \ . \label{eq:pbcx}
\end{align}
The correlations gathered individually in each measurement device can be directly obtained through marginalization. At this point, it is important to distinguish between different types of events in state discrimination. We say that a measurement successfully identified a preparation $x$ if the measurement outcome is equal to $x$. The average rate of occurrence of successful outcomes is denoted $p_s$. Similarly, if the measurement outcome is inside the alphabet of preparations $X$ but not equal to $x$, we say that the measurement performed an error, and we call $p_e$ the error probability. Additionally, measurement devices may lead to inconclusive events either through purposeful design or experimental imperfections. We label by $b,c=\varnothing$ any measurement outcome that is inconclusive---i.e.~is not in the alphabet of preparations $X$---and $p_{\varnothing}$ their outcome rate.

Moreover, post-measurement states cannot be arbitrarily chosen. For instance, consider that Bob employs a measurement strategy in a two-qubit state discrimination setting such that the error probability $p_e$ is null---also known as unambiguous state discrimination. During the rounds that his device returns conclusive results, i.e.~either $b=0$ or $b=1$, he will be able to send the original states $\rho_x$ to Charlie. However, if Bob's device returns an inconclusive event $b=\varnothing$, he will not know what state was produced and thus, will have to randomly choose to send either $\rho_0$ or $\rho_1$ to Charlie. It turns out that the optimal strategy for Bob and Charlie to optimally discriminate Alice's preparations with unambiguous qubit-state discrimination is to program Bob's instrument to produce two pure states $\ket{\phi_x}$ with an overlap lower-bounded by $\braket{\psi_0}{\psi_1}/p_{\varnothing}^{B}$ regardless of his measurement outcome \cite{bergou2013}, for $p_{\varnothing}^{B}$ being the probability of inconclusive events in Bob's device. A generalization with mixed states has also been studied in Ref.~\cite{hanwool2024}, where the optimal measurement choice becomes a maximum confidence measurement. The confidence $C_x$ that a receiver has in correctly discriminating a particular state is defined as the conditional probability that the state preparation was $x$ given that the observed measurement outcome is also $x$ \cite{croke2006}. It can be expressed through Bayes' rule as
\begin{align}
    C_x := \frac{p_x \Tr\left[\rho_x M_x\right]}{\sum_{x'} p_{x'} \Tr\left[\rho_{x'} M_x\right]} \ .
\end{align}
In the sequential maximum confidence scenario, optimality refers to cases where the same maximum confidence is shared by all participating parties. Maximum confidence measurements---i.e.~the measurements that yield the maximum value of confidence---have been widely studied in the context of qubit state discrimination (see Refs.~\cite{barnett2008,hanwool2022}). Additionally, it has been shown that they reproduce extremal prepare-and-measure correlations making them strong candidates for randomness certification \cite{carceller2024}. Maximum confidence measurements reduce to well-known state discrimination cases in extreme conditions. Concretely, on the one hand, if the probability of inconclusive events is null, maximum confidence measurements reduce to optimal minimum error state discrimination measurements \cite{bae2013}. On the other hand, if all state preparations are linearly independent, maximum confidence measurements reduce to optimal unambiguous state discrimination measurements \cite{ivanovic1987,dieks1988,peres1988}, with $C_x=1$, $\forall x$. Moreover, similarly as in a sequential unambiguous state discrimination scheme, Bob's post-measurement states cannot contain unlimited information about Alice's inputs $x$. Namely, as we show in \appref{app:seq_MCMs}, if Alice's preparations are qubits and initially contained a pure part with well defined overlaps $|\braket{\psi_0}{\psi_1}|=\delta$ and a  white noise component weighted by $1-r$, his post-measurement states can be chosen to be of the form $\sigma_x = t \ketbra*{\phi_x}{\phi_x} + \frac{1-t}{2}\mathds{1}$ with
\begin{align} \label{eq:t_fixed_main}
    t = r\sqrt{\frac{1-\delta^2+(1-r^2)\frac{\delta^2}{(p_{\varnothing}^{B})^2}}{1-r^2\delta^2}} \ .
\end{align}
and $|\braket{\phi_0}{\phi_1}|=\frac{r\delta}{p_{\varnothing}^{B}t}$. With this configuration, both Bob and Charlie are able to reach the same maximum confidence in discriminating $x$. In what follows, we investigate the certifiable randomness in specific cases with perfect knowledge of all state preparations, including the post-measurement states. Later, we generalize our findings to a case where only the overlaps of Alice's pure state preparations are known to be bounded, and Bob's post-measurement states are completely uncharacterised.

\section{Sequential randomness certification} \label{sec:seq_qrng}

The goal of Alice, Bob and Charlie is to be able to produce measurement outcomes that are as unpredictable as possible for an external malicious party. To this end, Bob and Charlie can perform maximum confidence measurements, each yielding an observable confidence $C^{\rm B}$ and $C^{\rm C}$, $\forall x$, and rates of inconclusive events $p_{\varnothing}^{\rm B}$ and  $p_{\varnothing}^{\rm C}$, respectively. To quantify how predictable those measurement outcomes are, we introduce the figure of an eavesdropper, or Eve, who has the goal of guessing those outcomes. We assume that, prior to the prepare-and-measure experiment, Eve has full access to the inner workings of the measurement devices and can pre-program the strategies $\lambda$ at will. Then, she hands the devices to all participating parties who privately select their respective inputs and observe their outcomes. Thus, the eavesdropping strategies $\lambda$ cannot depend on the inputs $x$. 

To this end, we present the tools based on semidefinite programming to compute bounds on the certifiable randomness. We consider two distinct cases. Firstly, a measurement-device-independent setting where all state preparations---even Bob's post-measurement states---are completely known. Secondly, we relax the assumption of known state preparations to a more semi-device-independent approach, where all devices are left uncharacterised and we only assume a bound on the pure state states prepared by Alice.

\subsection{Trusted state preparations} \label{sec:mdi}

We begin assuming complete knowledge of all state preparations, including Bob's post-measurement states, as sketched in \figref{fig:scenario}. Given a particular preparation $x^\ast$, we bound the certifiable randomness in Bob and Charlie's measurement outcomes individually. We find a bound on Eve's guessing probability $p_g$ which leads to a bound on the single-round min-entropy as $H_{\min}=-\log_2(p_g)$ through the leftover hash lemma \cite{renner2008,konig2009,tomamichel2017}. Specifically, given a set of $n$ known quantum states $\{\rho_x\}_{x=0}^{n-1}$, we maximize
\begin{align}\label{eq:pgB}
p_g^{B} :=& \sum_\lambda q(\lambda) \underset{b}{\max}\left\{\Tr\left[\rho_{x^\ast} M_{b}^{(\lambda)}\right]\right\} \ .
\end{align}
The maximization of $p_g^{B}$ runs over any possible POVM $\{M_{b}^{(\lambda)}\}$ and distributions $q(\lambda)$ such that Bob's confidence $C^{\rm B}$ and rate of inconclusive events $p_{\varnothing}^{\rm B}$ are reproduced. As we show in \appref{app:sdps}, we are able to cast this optimization as a semidefinite program \cite{vandenberghe1996}. Moreover, we reformulate the optimization problem using the dual semidefinite program, from which we find 
\begin{align} \label{eq:bound_pgB}
p_g^{B} \leq \ \min_{g,h,R,H^\lambda} \ g p_{\varnothing}^{\rm B}+hC^{\rm B}(1-p_{\varnothing}^{\rm B})+\Tr\left[R\right] \ , 
\end{align}
where $g$ and $h$ are arbitrary scalars restricted by
\begin{align}
\sum_{x} \rho_x \left(\delta_{b,\lambda}\delta_{x,x^\ast}-\frac{g}{n}\delta_{b,\varnothing}-\frac{h}{n}\delta_{b,x}\right)& \nonumber \\
+ H^{\lambda} - \frac{1}{d}\Tr\left[H^{\lambda}\right]\mathds{1} - R &\preceq 0 \ ,
\end{align}
for $R$ and $H^{\lambda}$ being arbitrary $d\times d$ Hermitian matrices. Formulating the semidefinite program in its dual form is relevant for computing randomness certification in real-world conditions. In experimental implementations, the observed frequencies of events only approximate the true underlying probability distribution. As a result, the estimated distribution may not correspond to any valid quantum realization, meaning that the associated constraints cannot be exactly satisfied. Since the dual optimal value always yields a lower bound on the primal solution, it will always find a solution even if the specified correlations---in this case in the form of observable confidence and rate of inconclusive events---are not physical.

We move on to computing the certifiable randomness produced in Charlie's device. After the first measurement on Alice's states, Bob is left with the post-measurement states $\{\sigma_x\}$ for which we have a full description. Concretely, let us assume that Alice's preparations were $\rho_x = r\ketbra*{\psi_x}{\psi_x} + (1-r)\frac{\mathds{1}}{d}$ where $d$ is the Hilbert space dimension, $r$ a noise parameter, and that the pure part of her states have a well-defined overlap $\abs{\braket{\psi_x}{\psi_{x'}}}=\delta$, $\forall x\neq x'$. Let Bob perform a maximum confidence measurement. Aligned with the framework from Ref.~\cite{hanwool2024}, if we want to maximize Charlie's confidence in the two-qubit state discrimination case, Bob's post-measurement states averaged over his measurement outcomes $b$ as in \eqref{eq:pm_states} will be given by $\sigma_x = t\ketbra*{\phi_x}{\phi_x} + (1-t)\frac{\mathds{1}}{d}$ with $t$ from \eqref{eq:t_fixed_main} and $\abs{\braket{\phi_x}{\phi_{x'}}}=\frac{r\delta}{p_{\varnothing}^{\rm B}t}$, $\forall x\neq x'$. 
These states are then sent to Charlie, who will perform a measurement described by the POVM $\{N_c^{(\lambda)}\}$ distributed according to $q(\lambda)$. For the concrete setting $x^\ast$, Eve's guessing probability is then given by
\begin{align}
p_g^{C} :=& \sum_\lambda q(\lambda) \underset{c}{\max}\left\{\Tr\left[\sigma_{x^\ast} N_{c}^{(\lambda)}\right]\right\} \label{eq:pgC} \ .
\end{align}
We aim to maximize $p_g^{C}$ over all possible POMVs implementable in Charlie's device, and over all distributions $q(\lambda)$. This optimization, however, is exactly the same scenario analogous to that from Alice-Bob prepare-and-measure case: instead of Alice preparing $\{\rho_x\}$ now Bob sends the states $\{\sigma_x\}$, and instead of Bob performing the measurement $\{M_{b}^{(\lambda)}\}$ now Charlie measures $\{N_{c}^{(\lambda)}\}$. We can thus use the exact same formalism and use the analogous form of the bound in \eqref{eq:bound_pgB}, which allows us to find a bound on the probability that Eve guesses Charlie's measurement outcome $p_g^{C}$, replacing $\rho_x\leftrightarrow\sigma_x$ and $M_b^\lambda\leftrightarrow N_c^\lambda$.

The proposed framework with trusted post-measurement states allows us to compute bounds on the randomness individually in Bob and Charlie's measurement outcomes using a single semidefinite program for each party. The underlying assumptions, although conveniently practical, are rather strong. For example, although Bob and Charlie's measurements are completely uncharacterised, Bob's post-measurement states $\{\sigma_x\}$ are assumed to be perfectly known and fixed over each prepare-and-measure round. If that were not true, Charlie's certifiable randomness would be drastically reduced, as Eve would have more freedom available to increase her guessing probability. More importantly, assuming a fixed form of post-measurement states over the rounds significantly limits potential adaptive eavesdropping strategies. Using adaptive attacking strategies, Eve can gather knowledge from attacks in Bob's device from previous rounds to improve her guess of $b$ in future rounds. Similarly, Eve can gather statistical information from previous attacks to both Bob and Charlie to improve her guess of $c$ in future rounds. However, if Bob's post-measurement states remain fixed and constant over the rounds, Eve cannot fully unleash her adaptive potential. That is because, after Bob's device employs measurement strategy $\lambda$, the trusted post-measurement states will be independent of $\lambda$ and thus, any different attack Eve could have performed in Bob will not influence her attack to Charlie. In consequence, the statistical behavior of Charlie's outcome will be independent of any knowledge she could have gathered during her attack in Bob and thus, Eve cannot employ a fully adaptive attacking strategy. This rules out the applicability of entropy accumulation methods \cite{metger2022}, which are becoming the standard approach in quantum cryptography (see e.g.~Refs.~\cite{wang2023,bhavsar2023,kamin2025}).

\subsection{Untrusted state preparations} 

\begin{figure}
	\centering
	\includegraphics[width=0.45\textwidth]{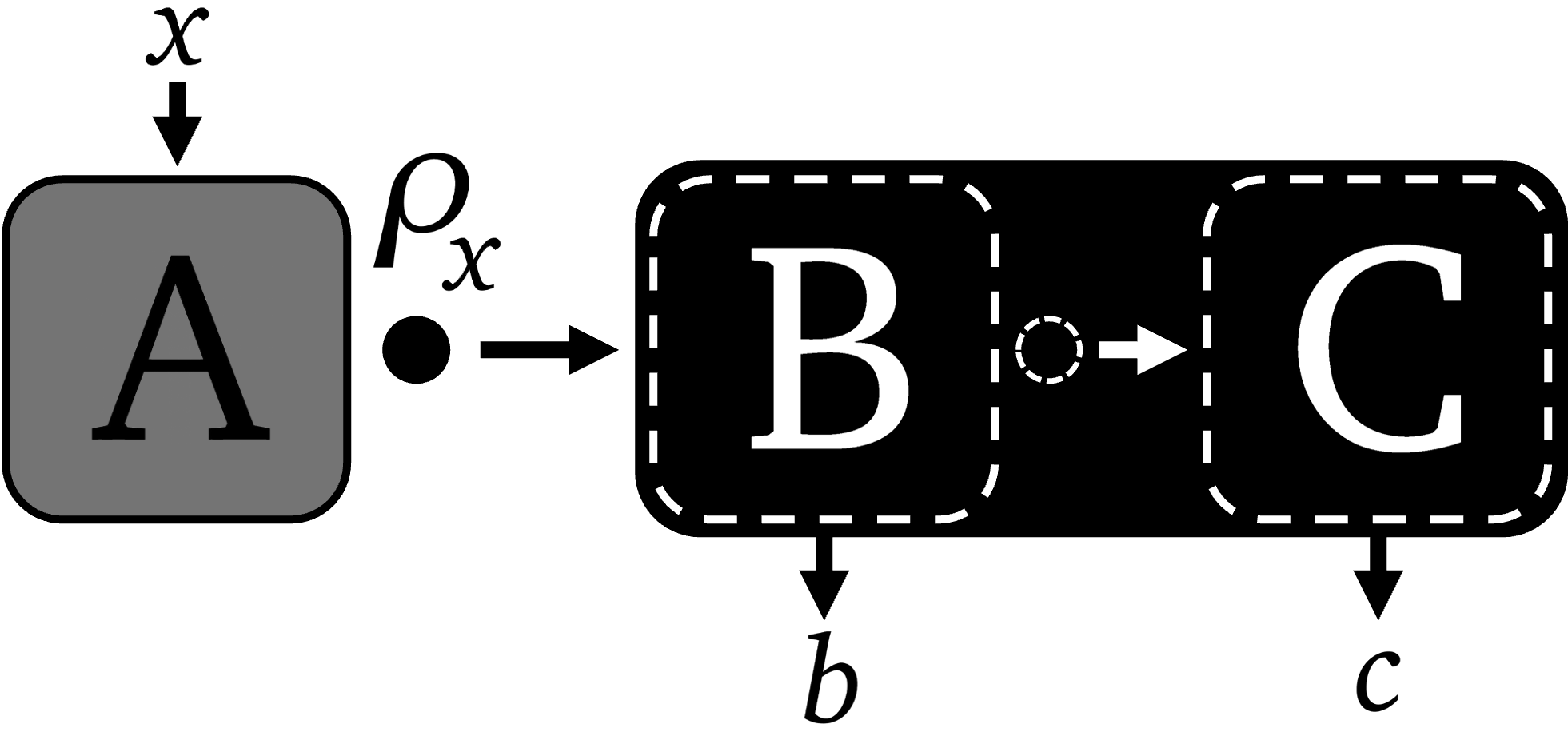}
  \caption{\textit{Semi-device-independent framework.} We assume that Alice's preparations are pure with bounded pair-wise overlaps. All operations performed after the preparation process are completely uncharacterised.}
 \label{fig:scenario2}
\end{figure}

In an attempt to relax the assumption of perfect knowledge in all state preparations, we introduce a more general framework as illustrated in \figref{fig:scenario2}. We remove any assumption on the structure of the post-measurement states and any intermediate operation between Bob and Charlie. Moreover, we assume that Alice's pure state preparations are only partially characterized through a lower-bound on their pair-wise overlap, i.e.~$\abs{\braket{\psi_x}{\psi_{x'}}}\geq \delta$.

In order to remove any imposed knowledge on Bob's post-measurement states, we collect any measurement and state updates he might perform in his device, together with Charlie's measurements, into the operator $G_{b,c}^{(\lambda)}:=K_b^{(\lambda)}N_{c}^{(\lambda)}K_b^{(\lambda)\dagger}$. This new operator constitutes a well-defined POVM. Namely, by definition, it is positive semidefinite $G_{b,c}\succeq 0$ $\forall b,c$ and complete $\sum_{bc} G_{b,c}=\mathds{1}$. Moreover, the observable correlations can be retrieved directly through the born rule $p(b,c|x)=\Tr\left[\rho_x G_{b,c}\right]$. It is important to remark that the sequential nature of the events $b$ and $c$ is not imprinted itself into the operator $G_{b,c}$. It is through the correlations---or more generally in any observable witness such as the confidence---, that the sequential nature of the operations embedded in $G_{b,c}$ must be imposed. 

Armed with the new compact definition of all sequential measurements in the operator $G$, we now present the methods to compute the certifiable randomness. We begin computing bounds on the min-entropy, and continue bounding Shannon entropy which we use to certify the randomness in non-independent and identically distributed (i.i.d.) prepare-and-measure rounds subject to adaptive attacking strategies.

\subsubsection{I.i.d.~rounds: min-entropy}

With the new operator $G_{b,c}$ describing any operation performed in Bob and Charlie's devices, we begin computing the min-entropy to quantify the single-shot certifiable randomness in Charlie's device (the certifiable randomness in Bob's device can be directly computed with the bound in \eqref{eq:bound_pgB}). To do so, we assume that we are given the observable confidences $C^{\rm B}$ and $C^{\rm C}$ and rates of inconclusive events $p_{\varnothing}^{\rm B}$ and $p_{\varnothing}^{\rm C}$ in Bob and Charlie's devices respectively. As explained in \secref{sec:mdi}, a lower-bound on the min-entropy can be found directly through upper-bounding the probability that Eve guesses the measurement outcome, conditioned on her having access to classical side information. Here, Eve's guessing probability of Charlie's outcome can be written as
\begin{align}\label{eq:pgC_sdi}
p_g^{\rm C} :=& \sum_b \sum_\lambda q(\lambda) \underset{c}{\max}\left\{\Tr\left[\rho_{x^\ast} G_{b,c}^{(\lambda)}\right]\right\}  \ .
\end{align}
The maximization of $p_g^{\rm C}$ can be formulated as a semidefinite program, as shown in \appref{app:sdps}. We further derive its dual form yielding the bound on Eve's guessing probability,
\begin{align}
p_g^{C} \leq \min_{\smash[b]{\substack{g^{\rm B},h^{\rm B},F^{\lambda_1,\lambda_2}\\ g^{\rm C},h^{\rm C},R}}}  \quad &g^{\rm B} p_{\varnothing}^{\rm B}+h^{\rm B} C^{\rm B}(1-p_{\varnothing}^{\rm B}) \label{eq:pgC_sdi_dual} \\ 
+&g^{\rm C} p_{\varnothing}^{\rm C}+h^{\rm C} C^{\rm C}(1-p_{\varnothing}^{\rm C})+\Tr\left[R\right] \ , \nonumber
\end{align}
where $g^{\rm B}$, $g^{\rm C}$, $h^{\rm B}$ and $h^{\rm C}$ are arbitrary scalars restricted by
\begin{align}
\!\sum_{x} \!\rho_x \!&\left(\!\delta_{c,\lambda}\delta_{x,x^\ast}\!-\!\frac{g^{\rm B}}{n}\delta_{b,\varnothing}\!-\!\frac{h^{\rm B}}{n}\delta_{b,x}\!-\!\frac{g^{\rm C}}{n}\delta_{c,\varnothing}\!-\!\frac{h^{\rm C}}{n}\delta_{c,x}\!\right) \nonumber \\
+& F^{\lambda} - \frac{1}{2}\Tr\left[F^{\lambda}\right]\mathds{1} - R \preceq 0 \ ,
\end{align}
with $R$ and $F^{\lambda}$ being arbitrary $d\times d$ Hermitian matrices.

By defining the new operator $G_{b,c}$ in the semi-device-independent framework, we can now also compute a bound on the certifiable randomness jointly contained in both Bob and Charlie's measurement outcomes. To do so, we first write the probability that Eve guesses both Bob and Charlie's outcomes as
\begin{align}
p_g^{BC} :=& \sum_\lambda q(\lambda) \underset{b,c}{\max}\left\{\Tr\left[\rho_{x^\ast} G_{b,c}^{(\lambda)}\right]\right\} \label{eq:pgBC} \ .
\end{align}
We maximize $p_g^{BC}$ for any set of pure states $\{\rho_x\}$ with pair-wise overlaps lower-bounded by $\delta$, and over all possible implementable $G_{b,c}^{(\lambda)}$, distributed according to $q(\lambda)$, such that the observable confidences $C^{\rm B}$ and $C^{\rm C}$, and rates of inconclusive events $p_{\varnothing}^{\rm B}$ and $p_{\varnothing}^{\rm C}$ are reproduced. As we show in detail in \appref{app:sdps}, we can cast this optimization problem as a semidefinite program, and subsequently find its dual formulation which will lead to $p_g^{\rm BC}$ be bounded by the exact same bound in \eqref{eq:pgC_sdi_dual}, but now $g^{\rm B}$, $g^{\rm C}$, $h^{\rm B}$ and $h^{\rm C}$ are arbitrary scalars subject to
\begin{align}
\!\sum_{x} \!\rho_x \!&\left(\!\delta_{b,\lambda_1}\!\delta_{c,\lambda_2}\!\delta_{x,x^\ast}\!-\!\frac{g^{\rm B}}{n}\delta_{b,\varnothing}\!-\!\frac{h^{\rm B}}{n}\delta_{b,x}\!-\!\frac{g^{\rm C}}{n}\delta_{c,\varnothing}\!-\!\frac{h^{\rm C}}{n}\delta_{c,x}\!\right) \nonumber \\
+& F^{\lambda_1 \lambda_2} - \frac{1}{2}\Tr\left[F^{\lambda_1 \lambda_2}\right]\mathds{1} - R \preceq 0 \ ,
\end{align}
where now we split the tuple $\lambda=(\lambda_1,\lambda_2)$ for each measuring party, with $R$ and $F^{\lambda_1 \lambda_2}$ being arbitrary $d\times d$ Hermitian matrices.

\subsubsection{Non-i.i.d.~rounds: Shannon entropy}

The semi-device-independent approach by compacting all sequential measurements in a single uncharacterized operation is now resilient to non-i.i.d.~attacks. Namely, Eve is allowed to adapt the attacking strategy during each round based on accumulated knowledge from previous rounds. This was not conceived in the measurement-device independent setting, which had an i.i.d.~component because post-measurement states were fixed on every round of the experiment. The channel connecting Bob and Charlie, however, must remain secured, i.e.~we assume that Eve cannot intercept Bob's post-measurement state in each round. This an assumption that, combined with classical-side-information, has been considered in the vast majority of previous semi-device-independent approaches \cite{vanhimbeeck2019,avesani2021,carceller2025photon,carceller2025improving}. 
Nevertheless, no knowledge on either the channel connecting both measurement devices, nor on the measurements themselves, is assumed. This makes the framework applicable for entropy accumulation theorems in prepare-and-measure scenarios \cite{metger2022}. The entropy accumulation theorem establishes that the total amount of entropy per-round accumulated after $m$ prepare-and-measure rounds assuming that Eve can perform adaptive attacking strategies over the rounds is bounded by
\begin{align}
    \frac{1}{m}H^{\varepsilon}_{\min}(B^m|E^m) \geq \min f_{min} - \mathcal{O}\left(\frac{1}{\sqrt{m}}\right) \ .
\end{align}
Here $H^{\varepsilon}_{\min}(B^m|E^m)$ is the so-called \textit{smooth min-entropy} accumulated over $m$ prepare-and-measure rounds \cite{renner2008}, and $f_{min}$ is an affine function that lower-bounds the Shannon entropy (also known as \textit{min-tradeoff function}). This quantity is essential to treat finite size effects in non-i.i.d.~prepare-and-measure rounds. In what follows, we introduce a method to find valid min-tradeoff functions by lower-bounding the Shannon entropy based on the recent works from Refs.~\cite{carceller2025photon,carceller2025improving}. Concretely, let us begin considering Bob's system. The Shannon entropy in Bob's system $B$ given a particular choice of preparation $x^\ast \in X$ and conditioned on Eve's system $E$ and on the classical side information stored in the shared randomness $\lambda$ can be written as
\begin{align}\label{eq:HB}
H(B|E) = -\sum_\lambda q(\lambda)\sum_{b}p_\lambda(b|x^\ast)\log_2p_\lambda(b|x^\ast) \ ,
\end{align}
for $p_\lambda(b|x)=\sum_c \Tr\left[\rho_{x} G_{b,c}^{(\lambda)}\right]$. As we show in \appref{app:sdps}, we are able to compute an affine min-tradeoff function $f_{min}^{\rm B}\leq H(B|E)$ by expressing the Shannon entropy through the Gauss-Radau quadrature following the ideas from Ref.~\cite{brown2024}. One can then cast the optimisation as a semidefinite program relaxation and find the subsequent dual form. This provides us with the min-tradeoff function
\begin{align}
f_{min}^{\rm B} = c_m &- g p_{\varnothing}^{\rm B}-hC^{\rm B}(1-p_{\varnothing}^{\rm B})-\Tr\left[R\right] , \label{eq:mintradeofB}
\end{align}
for $c_m=\sum_{i=1}^{m-1}\frac{w_{i}}{t_{i}\ln{2}}$, with $t_i$ and $w_i$ being the nodes and weights from the Gauss-Radau quadrature which can be obtained through a simple algorithm \cite{golub1973}. Here  $g$, $g^{\rm C}$, $h^{\rm B}$ and $h^{\rm C}$ are arbitrary scalars and $R$ arbitrary $d\times d$ Hermitian matrices subject to
\begin{align}
&\sum_{ij} D_{b}^{ij} = \sum_{x}\rho_x \left(\frac{g}{n}\delta_{b,\varnothing}+\frac{h}{n}\delta_{b,x}\right) + R \nonumber   \\
&F_{b}^{ij} + F_{b}^{ij\dagger} = 2\tau_i\rho_{x^\ast}\delta_{j,b} + Q_1^{ij} - \frac{1}{d}\Tr\left[Q_1^{ij}\right]\mathds{1} \\
&L_{b}^{ij}\!=\!\tau_i\rho_{x^\ast}\!\left[\left(1\!-\!t_i\right)\delta_{j,b}\!+\!t_i\right]\! +\! Q_2^{ij}\! -\! \frac{1}{d}\Tr\left[Q_2^{ij}\right]\mathds{1}   \nonumber \ .
\end{align}
Here $Q_1^{ij}$, $Q_2^{ij}$, $D_{b}^{ij}$, $L_{b}^{ij}$ and $F_{b}^{ij}$ are arbitrary $d\times d$ Hermitian matrices, the last three constrained to fulfill the positivity of the following block-matrix
\begin{align}
\begin{pmatrix}
        D_{bc}^{ij} & F_{b}^{ij} \\
        F_{b}^{ij\dagger} & L_{b}^{ij}
    \end{pmatrix} \succeq 0  \ .
\end{align} 
Given a set of variables that fulfill the specified constraints, the min-tradeoff function in \eqref{eq:mintradeofB} represents a lower bound on the Shannon entropy valid for any observable confidence $C^{\rm B}$ and rates of inconclusive events $p_{\varnothing}^{\rm B}$ in Bob's device, respectively.

\begin{figure*}
\includegraphics[width=1.0\textwidth]{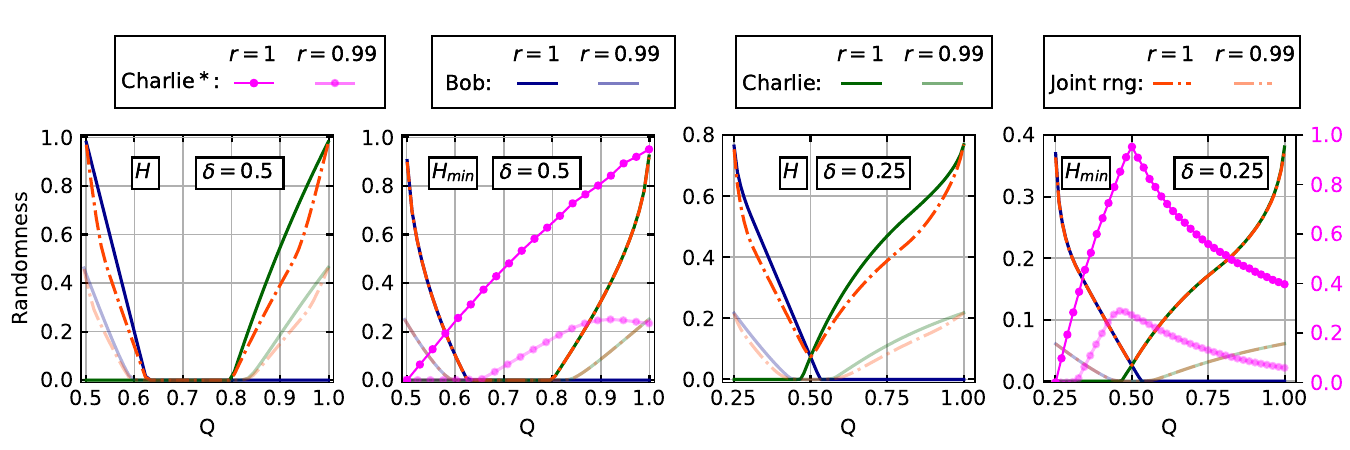}
\caption{Shannon and min-entropies certified in two sequential maximum confidence measurements. Alice prepares two pure states with $\left|\braket{\psi_0|\psi_{1}}\right|\geq \delta$. Bob and Charlie perform maximum confidence measurement with white noise component $1-r$, and rates of inconclusive outcomes $Q$ and $r\delta/Q$ respectively. We also show the certifiable min-entropy in Charlie's device in the measurement-device independent setting (Charlie$^\ast$).}
\label{fig:results}
\end{figure*}

We continue computing a bound on the certifiable Shannon entropy in Charlie's device, and in both Bob and Charlie's devices jointly. Respectively, the Shannon entropy in Charlie's system $C$ and Bob $+$ Charlie's systems $BC$ conditioned on a particular input $x^\ast \in X$ and conditioned on Eve's system $E$ and on the classical side information stored in the shared randomness $\lambda$ can be written as
\begin{align}
H(C|E) &= -\sum_\lambda q(\lambda)\sum_{c}p_\lambda(c|x^\ast)\log_2p_\lambda(c|x^\ast) \label{eq:HC} \\
H(BC|E) &= -\sum_\lambda q(\lambda)\!\sum_{b,c}p_\lambda(b,c|x^\ast)\log_2p_\lambda(b,c|x^\ast) \label{eq:HBC} ,
\end{align}
for $p_\lambda(b,c|x)=\Tr\left[\rho_{x} G_{b,c}^{(\lambda)}\right]$ and $p_\lambda(c|x)=\sum_b p_\lambda(b,c|x)$. Similalry as in Bob's case, in \appref{app:sdps} we show that we are able to find an affine min-tradeoff function $f_{min}^{C}\leq H(C|E)$ and $f_{min}^{BC}\leq H(BC|E)$ by expressing the optimizations in a semidefinite program and deriving their dual form. This provides us with the exact same bound in both cases $f_{min}$, i.e.
\begin{align}
f_{min} = c_m &- g^{\rm B} p_{\varnothing}^{\rm B}-h^{\rm B}C^{\rm B}(1-p_{\varnothing}^{\rm B}) \nonumber \\
&- g^{\rm C} p_{\varnothing}^{\rm C}-h^{\rm C}C^{\rm C}(1-p_{\varnothing}^{\rm C}) - \Tr\left[R\right] \label{eq:mintradeof}
\end{align}
Depending on whether we are interested in bounding the Shannon entropy in solely Charlie's $C$ or the joint $BC$ systems, the scalars $g^{\rm B}$, $g^{\rm C}$, $h^{\rm B}$ and $h^{\rm C}$ and $d\times d$ Hermitian matrices $R$ will be subject to different restrictions. For $f_{min}\leq H(C|E)$,
\begin{align}
&\sum_{ij} D_{bc}^{ij} = \sum_{x}\rho_x\left(\frac{g^{\rm B}}{n}\delta_{b,\varnothing}+\frac{h^{\rm B}}{n}\delta_{b,x} + \frac{g^{\rm C}}{n}\delta_{c,\varnothing}+\frac{h^{\rm C}}{n}\delta_{c,x}\right) + R \nonumber   \\
& F_{bc}^{ij} + F_{bc}^{ij\dagger} = 2\tau_i\rho_{x^\ast}\delta_{j,c} + Q_1^{ij} - \frac{1}{d}\Tr\left[Q_1^{ij}\right]\mathds{1} \\
&L_{bc}^{ij}\!=\!\tau_i\rho_{x^\ast}\!\left[\left(1\!-\!t_i\right)\delta_{j,c}\!+\!t_i\right]\! +\! Q_2^{ij}\! -\! \frac{1}{d}\Tr\left[Q_2^{ij}\right]\mathds{1}  ,  \nonumber
\end{align}
where $Q_1^{ij}$, $Q_2^{ij}$, $D_{bc}^{ij}$, $L_{bc}^{ij}$ and $F_{bc}^{ij}$ are arbitrary $d\times d$ Hermitian matrices satisfying the positivity constraint
\begin{align}
\begin{pmatrix}
        D_{bc}^{ij} & F_{bc}^{ij} \\
        F_{bc}^{ij\dagger} & L_{bc}^{ij}
    \end{pmatrix} \succeq 0  \ .
\end{align} 
On the other hand, for $f_{min}\leq H(BC|E)$,
\begin{align}
&\!\!\!\sum_{ijk} D_{bc}^{ijk}\! =\! \sum_{x}\rho_x\left(\!\frac{g^{\rm B}}{n}\delta_{b,\varnothing}\!+\!\frac{h^{\rm B}}{n}\delta_{b,x} \!+\! \frac{g^{\rm C}}{n}\delta_{c,\varnothing}\!+\!\frac{h^{\rm C}}{n}\delta_{c,x}\!\right)\! +\! R \nonumber   \\
&F_{bc}^{ijk} + F_{bc}^{ijk\dagger} = 2\tau_i\rho_{x^\ast}\delta_{j,b}\delta_{k,c} + Q_1^{ijk} - \frac{1}{d}\Tr\left[Q_1^{ijk}\right]\mathds{1} \\
&L_{bc}^{ijk}\!=\!\tau_i\rho_{x^\ast}\!\left[\left(1\!-\!t_i\right)\delta_{j,b}\delta_{k,c}\!+\!t_i\right]\! +\! Q_2^{ijk}\! -\! \frac{1}{d}\Tr\left[Q_2^{ijk}\right]\mathds{1}   \nonumber
\end{align}
Here $Q_1^{ijk}$, $Q_2^{ijk}$, $D_{bc}^{ijk}$, $L_{bc}^{ijk}$ and $F_{bc}^{ijk}$ are arbitrary $d\times d$ Hermitian matrices, the last three constrained to fulfill the positivity of the block-matrix
\begin{align}
\begin{pmatrix}
        D_{bc}^{ijk} & F_{bc}^{ijk} \\
        F_{bc}^{ijk\dagger} & L_{bc}^{ijk}
    \end{pmatrix} \succeq 0  \ .
\end{align} 
Given a set of variables that fulfill the specified constraints, the min-tradeoff function in \eqref{eq:mintradeof} represents an affine lower bound on the Shannon entropy valid for any observable confidences $C^{\rm B}$, $C^{\rm C}$ and rates of inconclusive events $p_{\varnothing}^{\rm B}$, $p_{\varnothing}^{\rm C}$ in Bob and Charlie's devices, respectively.

\section{Results} \label{sec:res}

To illustrate our methods, we consider a simple two-state discrimination scenario with two measurements (Bob and Charlie) performed sequentially. We test our approach when Bob and Charlie perform maximum confidence measurements, and Bob's performance is dictated by a limit in his rate of inconclusive events.

First, we begin bounding the certifiable confidence which has been recently studied in the context of sequential state discrimination \cite{hanwool2024}. We assume that Alice is able to prepare a pair of pure states $\rho_x=\ketbra*{\psi_x}{\psi_x}$ for $x=0,1$, that have an overlap that satisfies the bound $\abs{\braket{\psi_0}{\psi_1}} \geq \delta$. Next, on the one hand, Bob's performance is parametrized through a bound on his rate of inconclusive events $p_{\varnothing}^{B}\geq Q$ which we range from the minimum value allowed by noiseless maximum confidence measurements (i.e.~unambiguous state discrimination) and the algebraic maximum, i.e.~$\delta\leq Q\leq 1$. 
Additionally, we compute a bound on the certifiable confidence in Bob's side $C^{B}$ considering white noise. Namely, although Bob receives noiseless pure states from Alice, the correlations he gathers in his device are sub-optimal and compatible with those he would have gathered with noisy states of the from $r\ketbra*{\psi_x}{\psi_x} + \frac{1-r}{d}\mathds{1}$. On the other hand, we compute a bound on Charlie's maximum confidence considering that Bob's device reproduces the named bound on the inconclusive rate $Q$ and observed confidence $C^{B}$. We fix Charlie's rate of inconclusive events to reach the optimal value $p_{\varnothing}^{C}=r\delta / Q$, i.e.~the minimum allowed by optimal maximum confidence measurements. This follows from the fact that we assume both parties attain the maximum confidence, which will be equal as shown earlier.

Next, we use the observable confidences and resulting rates of inconclusive events to bound the certifiable randomness in Bob and Charlie's measurement outcomes when $\delta=0.5$ and $\delta=0.25$. In \figref{fig:results} we show the certifiable randomness bounds. Concretely, we consider the approach in which all state preparations, including the post-measurement states, are fully characterised, and compute a bound on the min-entropy ($H_{\min}$) in Bob and Charlie's devices through the guessing probabilities in \eqref{eq:pgB} (blue line labeled Bob) and \eqref{eq:pgC} (pink-dotted line labeled Charlie$^\ast$) respectively. We then move on with the second approach, with uncharacterised post-measurement states, and compute a bound on the i.i.d.~randomness with the min-entropies computed through the guessing probabilities in \eqref{eq:pgC} (green line labeled Charlie) and  \eqref{eq:pgBC} (red line labeled Joint rng) in Charlie's and Bob$+$Charlie's systems respectively (Bob's randomness in this case is equivalent to the blue line labeled Bob). Finally, we consider the non-i.i.d.~case and compute a bound on the Shannon entropy ($H$) in Bob's system from \eqref{eq:HB}, in Charlie's system from \eqref{eq:HC}, and in the joint Bob$+$Charlie's systems through \eqref{eq:HBC}.

Looking at the first approach with fully trusted preparations, we see that Charlie's certifiable randomness peaks at inconclusive rates $Q=\frac{r\delta}{2}$. This happens because we are imposing a highly restrictive assumption in the form of Bob's post-measurement states. Namely, these are constrained to be of the form $\sigma_x = t\ketbra*{\phi_x}{\phi_x} + (1-t)\frac{\mathds{1}}{d}$ with $t$ from \eqref{eq:t_fixed_main} and $|\braket{\phi_x}{\phi_{x'}}|=\frac{r\delta}{p_{\varnothing}^{\rm B}t}$. Then, Charlie's maximum confidence measurement with an optimal rate of inconclusive events $p_{\varnothing}^{\rm C}=\frac{r\delta}{Q}$ yields maximal randomness in the measurement outcome for $p_{\varnothing}^{\rm C}=\frac{1}{2}$. This high randomness yield reflects the strength of the assumptions placed in our first approach, i.e.~assuming perfect knowledge on Bob's post-measurement states.

Through the second approach (untrusted post-measurement states), we observe that the certifiable randomness in Bob's device dies out when the rate of inconclusive outcomes reaches a critical value $Q_{\rm crit}^{\rm B}=\frac{1+r^2\delta^2}{2}$. The fact that for certain rates $Q$ there exist no certifiable randomness indicates that the observable confidence $C^{B}$ and rate $p_{\varnothing}^{B}$ can be reproduced through predefined deterministic measurement schemes. This is indeed what is happening here, and we will show it through a simple construction (see \appref{app:inc_lims} for more details). Let $\ket{\varphi_x}$ be the set of complementary states of Alice's preparations such that $\rho_0=C^{B}\ketbra*{\varphi_0}{\varphi_0}+(1-C^{B})\ketbra*{\varphi_1}{\varphi_1}$ and $\rho_1=(1-C^{B})\ketbra*{\varphi_0}{\varphi_0}+C^{B}\ketbra*{\varphi_1}{\varphi_1}$, with $\abs{\braket{\varphi_x}{\varphi_{x'}}}=r\delta$, $\forall x\neq x'$. Let the measurement strategy in Bob's device $\{M_b^\lambda\}$ be determined by the dichotomic hidden variable $\lambda\in\{0,1\}$ equally distributed according to $q(\lambda)=\frac{1}{2}$. If $\lambda=0$, let Bob's measurement scheme be $M_1^0=(2-c)\ketbra*{\varphi_0^\perp}{\varphi_0^\perp}$ and $M_{\varnothing}^0=(2-c)\ketbra*{\varphi_0}{\varphi_0}+(c-1)\mathds{1}$ and never produce the outcome $b=0$. Otherwise, if $\lambda=1$ let Bob never produce outcome $b=1$ and $M_1^0=(2-c)\ketbra*{\varphi_1^\perp}{\varphi_1^\perp}$ and $M_{\varnothing}^1=(2-c)\ketbra*{\varphi_1}{\varphi_1}+(c-1)\mathds{1}$. With this deterministic measurement strategy, the averaged rate of inconclusive events is $p_{\varnothing}=\sum_\lambda q(\lambda) p_{\varnothing}^\lambda = r^2 \delta^2 + \frac{1-r^2 \delta^2}{2}c$, which is limited by $\frac{1+r^2\delta^2}{2}\leq p_{\varnothing}\leq 1$ since $1\leq c \leq 2$. Similarly for Charlie, his certifiable randomness is null until Bob's rate of inconclusive events reaches a second critical value $Q_{\rm crit}^{\rm C}=\frac{2r\delta}{1+r^2\delta^2}$. With maximum confidence discrimination therefore, no randomness can be sequentially certified simultaneously in Bob and Charlie's devices if Alice's pure state preparations have an overlap $\delta$ and white noise component $1-r$ that fulfill $r\delta\leq \Delta$, for $\Delta \simeq 0.2956$ (that is, when $Q_{\rm crit}^{\rm B}\leq Q_{\rm crit}^{\rm C}$). Otherwise, randomness can be simultaneously certified in both devices if $Q_{\rm crit}^{\rm C}\leq Q \leq Q_{\rm crit}^{\rm B}$. This insight indicates that there are some cases in which randomness can be simultaneously certified in two sequential measurements. In the following, we complete this analysis by extending these findings to an arbitrary number of sequential maximum confidence measurements.

\begin{figure}
	\centering
	\includegraphics[width=\columnwidth]{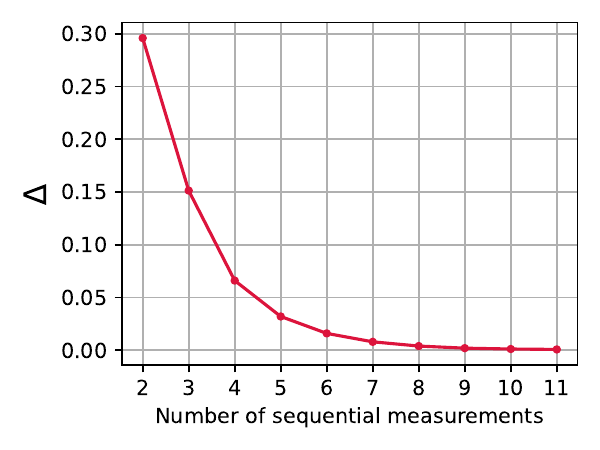}
  \caption{Numerical search of the bound $r\delta \leq \Delta$ at which randomness can be simultaneously certified in all $N$ sequential maximum confidence measurements.}
 \label{fig:find_N}
\end{figure}

\section{Extension to $N$ sequential measurements} \label{sec:n_seqs}

For now we only considered the simplest case of two sequential maximum confidence measurements. Here, we now consider the extended scenario to an arbitrary number $N$ of sequential maximum confidence measurements. Let $\{S_1,S_2,\ldots,S_N\}$ label the systems of each measurement and let $p_{\varnothing}^{S_i}\geq Q_i$ be the rate of inconclusive events in each individual system. Again, consider that all the observable statistics are compatible with those reproduced by maximum confidence measurements on two state preparations of the form $r\ketbra*{\psi_x}{\psi_x} + \frac{1-r}{2}\mathds{1}$ with $\abs{\braket{\psi_0}{\psi_1}}\geq \delta$. Let the post-measurement states after the $i^{th}$ measurement be $t_i\ketbra*{\phi_x^i}{\phi_x^i} + \frac{1-t_i}{2}\mathds{1}$ with $\abs{\braket{\phi_0^i}{\phi_1^i}} = s_i$. To determine the criteria at which randomness can be simultaneously certified in all sequential measurements, we generalize the argument we employed in the two sequential measurement case to the $N$ case. That is, each measurement needs to reproduce a rate of inconclusive events below the threshold which can be reproduced by deterministic strategies. Thus, to certify randomness during the $i^{th}$ sequential measurement that receives states with a pure part with overlap $s_{i-1}$ and purity weight $t_{i-1}$, the condition
\begin{align} \label{eq:general_Qbound}
    Q_i \leq \frac{1+t_{i-1}^2 s_{i-1}^2}{2} = \frac{1}{2}\left(1+\frac{r^2\delta^2}{\prod_{j=0}^{i-1}Q_j^2}\right)
\end{align}
needs to be satisfied. In the equality we used the fact that if maximum confidence measurements are performed, the relation $t_{i} s_{i} = \frac{t_{i-1}s_{i-1}}{Q_{i}}$ has to hold, with $t_0=r$, $s_0=\delta$ corresponding to the initial state preparations and $Q_0=1$. Therefore, if the condition in \eqref{eq:general_Qbound} holds $\forall i$, randomness can be simultaneously certified in all outcomes of the sequential measurements. In analogy with the example shown in \figref{fig:results}, we take the case that the last measurement yields the minimum rate of inconclusive events compatible with a maximum confidence measurement, this way consuming all certifiable randomness left after all posterior measurements,i.e.~, $Q_N=\frac{r\delta}{\prod_{j=0}^{i-1}Q_j}$. In order to be able to certify randomness in all devices, the parameters $\delta$ and $r$ need to be the ones that satisfy the condition $Q_N\geq Q_i$, $\forall i$, given that $Q_i\leq Q_{i+1}$ according to \eqref{eq:general_Qbound}. We performed a numerical search for all $\delta$ and $r$ that satisfy that condition and gathered in the multiplication $r\delta$. In \figref{fig:find_N} we show the bound $r\delta \leq \Delta$ that needs to be satisfied in order to be able to sequentially certify randomness in $N$ consecutive maximum confidence measurements. In a nutshell, if the pure state preparations have an overlap lower-bounded by $\delta$, and the observable correlations are compatible with noisy states with $1-r$ white noise weight, a set of $n$ sequential measurements will be able to certify randomness if the product $r\delta$ is below the $n^{th}$ point in \figref{fig:find_N}. As the number of parties increases, the bound $\Delta$ rapidly decreases, to a point which it is practically impossible to certify randomness in all sequential measurements, making it pointless to add subsequent measurements.

\section{Discussion} \label{sec:disc}

We investigated the certification of randomness in sequential prepare-and-measure scenarios through maximum-confidence state discrimination. Our approach is semi-device-independent, where assumptions on devices can be significantly relaxed compared to more stringent device-independent approaches based on Bell scenarios. Centrally, we showed that quantum randomness, originating from the fundamental indistinguishability of non-orthogonal states, can be distributed and certified among multiple observers through maximum-confidence measurements. 

From a technical point of view, we introduced rigorous methods to quantify randomness in sequential prepare-and-measure scenarios with trusted state preparation (including post-measurement states) and uncharacterised measurements. By leveraging semidefinite programming techniques, we derived tight bounds on Eve’s guessing probability, which directly translate into lower-bounds on the certifiable min-entropy. Secondly, we generalized the sequential randomness certification framework to a setting where only the overlap of the (pure) state preparations is trusted. Here, the sequential structure was compactly encoded into effective operators, which allowed us to extend the certification methods to adaptive attack strategies and non-i.i.d.~scenarios, and enabled the use of entropy accumulation methods. Our methods rely on semidefinite programming duality, ensuring their feasibility even in the presence of experimental imperfections that may render the observable frequencies with no physical realization. This makes the approach immediately applicable to realistic scenarios, where noise and limited control over devices are unavoidable. 

From a foundational perspective, our analysis clarifies the subtle connection between quantum state discrimination and the distribution of randomness in sequential processes. Maximum-confidence measurements, in particular, emerge as strong candidates for certifying randomness, as they interpolate between well-known state discrimination strategies---in one extreme minimum error state discrimination and in the other unambiguous state discrimination---while reproducing extremal prepare-and-measure correlations. Deploying maximum confidence measurements in the sequential scenario reveals that randomness is not an isolated property of a single measurement but can be shared among multiple parties. Our analysis provided a sufficient criterion for simultaneously extracting randomness from a given number of sequential measurements, which only depends on the distinguishability of the initial state preparations.

The relevance of these findings extends well beyond the theoretical domain. Sequential prepare-and-measure implementations can be realized with relatively modest experimental resources, e.g.~on integrated photonic platforms supporting non-destructive quantum measurements \cite{solis2016,namkung2018,anwer2020}. Likewise, optical systems provide a natural setting for the implementation of maximum-confidence measurements \cite{mosley2006,steudle2011,gomez2022,melo2023}. In photonic platforms, energy-related observables such as photon-number correlations or constraints on the vacuum component arise naturally as measurable quantities, and there is a growing body of research proposing semi-device-independent frameworks based on energetic constraints \cite{vanhimbeeck2017,rusca2019,Tebyanian2021b,carceller2025depth}. The approach introduced here is not confined to the specific scenario analyzed in this work. It can be extended to semi-device-independent frameworks defined by energy constraints, by reformulating the corresponding entropy optimization problem through appropriate semidefinite programming relaxations \cite{tavakoli2024}.

Finally, our results demonstrate that quantum unpredictability can be sequentially distributed, even under partial trust assumptions. This naturally raises the question of whether such distributed unpredictability can be harnessed beyond randomness certification, e.g., in semi-device-independent quantum key distribution. A particularly intriguing direction is the exploration of multipartite cryptographic schemes, such as quantum secret sharing and third-man quantum cryptography protocols \cite{zukowski1998,chen2005}, or sequential variants of the BB84 protocol \cite{bennett2014}, where additional collaborative measurements performed after the first receiver could enhance key generation rates, or even enable secure key distribution among multiple parties.

\begin{acknowledgements}
    C.R.C. is supported by the Wenner-Gren Foundations. J.B.B. acknowledges the Danish National Research Foundation grant bigQ (DNRF 142). H.L acknowledges the financial support from the Business Finland project BEQAH. K.F. and J.B. are supported by NRF Korea (2021R1A2C2006309, 2022M1A3C2069728) and the IITP (RS-2023-00229524, RS-2025-02304540). 
\end{acknowledgements}

\section*{Code availability}

The codes used to generate the results in this paper are available in GitHub: \url{https://github.com/chalswater/Sequential-QRNG-with-MCM}.

\bibliography{seq_qrng}

\appendix

\begin{widetext}

\newpage

\section{Sequential maximum confidence measurements}
\label{app:seq_MCMs}

In this part of the supplemental material we show that, in a sequential prepare-and-measure scenario, Bob cannot perfectly encode information about the initial state preparations $x$ in his post-measurement states using maximum confidence measurements. We show that both Bob and Charile can obtain the same maximum confidence when they perform sequential measurements. 

Consider that Alice prepares two qubit states of the form
\begin{align}
\rho_x = r\ket{\psi_x}\bra{\psi_x} + \frac{1-r}{2}\mathds{1} \label{eq:statesapp}
\end{align}
with prior probability $\frac{1}{2}$, where $1-r$ is the amount of white noise introduced by Alice's device and $\left\{\ket{\psi_x}\right\}$ for $x=0,1$ are two pure states with a well defined overlap $\braket{\psi_0|\psi_1}=\delta$. We can parametrize these states as
\begin{align}
    \ket{\psi_x}=\sqrt{\frac{1+\delta}{2}}\ket{0}+(-1)^{x}\sqrt{\frac{1-\delta}{2}}\ket{1} \ .
\end{align}
Confidence is defined as a conditional probability that given a measurement outcome $x$ correctly identifies the preparation $x$. By using Bayes' rule, the maximum confidence of $x$ can be found by the following optimization
\begin{align}
C_x = \max_{M_x\geq 0}\frac{1}{2}\frac{\Tr\left[\rho_x M_x\right]}{\Tr\left[\rho M_x\right]} \ ,
\end{align}
where $\rho=\frac{1}{2}(\rho_0+\rho_1)$ and $\{M_b\}$ are the POVM elements describing Bob's measurement. 

Bob receives the states prepared by Alice and performs a maximum confidence measurement. Let us define the following pure states,
\begin{align}
    \ket{\varphi_{x}} =    \sqrt{ \frac{1+p\cos\theta }{{2}}}\ket{0} +(-1)^{x }\sqrt{ \frac{1-p\cos\theta}{{2}}}\ket{1}, x=0,1 \ . ~~~
\end{align}
As derived in \cite{hanwool2024}, a POVM defined as 
\begin{align}
   &M_0=c_0 \ket{\varphi_1^\perp}\bra{\varphi_1^\perp} \\
   &M_1=c_1\ket{\varphi_0^\perp}\bra{\varphi_0^\perp} \nonumber \\
   &M_\phi=I-M_0-M_1 \nonumber
\end{align}
is a maximum confidence measurement where $c_0,c_1 \geq 0$ are constants related to the inconclusive rate. This measurment yields the maximum confidence
\begin{align}
    C_x= \frac{1}{2}\left(1+\frac{r\sqrt{1-\delta^2}}{\sqrt{1-r^2\delta^2}}\right) \ ,
\end{align}
whcih we denote as $C$. Although the Kraus operators can be freely chosen such that $K_b^\dag K_b=M_b$, here we consider the particular choice 
\begin{align}
    &K_0=\sqrt{c_0} \ket{\xi_0} \bra{\varphi_1^\perp}\\
    &K_1=\sqrt{c_1} \ket{\xi_0} \bra{\varphi_1^\perp} \nonumber\\
    &K_\phi=\sqrt{a_0} \ket{\xi_0} \bra{\varphi_1^\perp}+\sqrt{a_1} \ket{\xi_1} \bra{\varphi_0^\perp} \, ,\nonumber
\end{align}
where $a_0, a_1 \geq 0$ are parameters that satisfy $\sum_{b} K_b^\dag K_b=\mathds{1}$.This particular Kraus operators allow Bob and Charlie attain the same maximum confidence $C$ as we will see in the following.

Note that the states in Eq. (\ref{eq:statesapp}) can be re-written in the following convenient notation
\begin{align}
\rho_0 &= C\ket{\varphi_0}\bra{\varphi_0} + (1-C)\ket{\varphi_1}\bra{\varphi_1} \\
\rho_1 &= (1-C)\ket{\varphi_0}\bra{\varphi_0} + C\ket{\varphi_1}\bra{\varphi_1} \nonumber
\end{align}
and these two pure states have an overlap $\braket{\varphi_0|\varphi_1}=r\delta$.
After performing a measurement, Bob spits the post-measurement states $\sigma_x =\sum_{b} K_b \rho_x K_b^\dag$, where
\begin{align}
\sigma_0 &= C\ket{\xi_0}\bra{\xi_0} + (1-C)\ket{\xi_1}\bra{\xi_1} \\
\sigma_1 &= (1-C)\ket{\xi_0}\bra{\xi_0} + C\ket{\xi_1}\bra{\xi_1} \nonumber \ .
\end{align}
Charlie's maximum confidence measurement $\{N_c\}$ is
\begin{align}
    &N_0=d_0 \ket{\xi_1^\perp}\bra{\xi_1^\perp}\\
    &N_1=d_1 \ket{\xi_0^\perp}\bra{\xi_0^\perp} \nonumber\\
    &N_\phi=\mathds{1}-N_0-N_1 \nonumber \ ,
\end{align}
where $d_0, d_1 \geq 0$ are constants related to the inconclusive rate of Charlie, and this measurement yields the confidence $C$ which is the same as Bob's confidence. 

In Ref. \cite{hanwool2024} it is shown that $\{\ket{\xi_x}\}$ are pure states with overlap $\langle \xi_0 | \xi_1 \rangle > \langle \varphi_0 | \varphi_1 \rangle$. In particular, if we impose a condition that $c_0=c_1$ and $a_0=a_1$, then
\begin{align}
    \langle \xi_0 | \xi_1 \rangle=\frac{\langle \varphi_0 | \varphi_1 \rangle}{p_\varnothing^B} =\frac{r \delta}{p_\varnothing^B}\ ,
\end{align}
where $p_\varnothing^B=\Tr[\rho M_\phi]$ is the inconclusive rate by Bob. 
Let us write the post-measurement states as
\begin{align} \label{eq:Bob_pmstates}
\sigma_x =  t \ket{\phi_x}\bra{\phi_x} + \frac{1-t}{2}\mathds{1} \ ,
\end{align}
where $1-t$ is the amount of white noise introduced by Bob's device and $\{\ket{\phi_x}\}$ are two pure states with an overlap $\braket{\phi_0|\phi_1}=s$.
Then, the pure states $\{ \ket{\xi_x}\}$ have an overlap $\braket{\xi_0|\xi_1}=ts$.

The relationship between the initial states $\rho_x$ and the post-measurement states $\sigma_x$ is parameterized by the overlaps $\delta$ and $s$ and white noise parameters $r$ and $t$. For instance, take both descriptions from the post-measurement states. Their purity is given by
\begin{align}\label{eq:purity_pmstates_1}
    \Tr\left[\sigma_x^2\right]=\frac{1+t^2}{2}=1-2C(1-C)\left(1-|\braket{\xi_0|\xi_1}|^2\right) \ .
\end{align}
Since $\langle \xi_0 | \xi_1 \rangle = \frac{r\delta}{p_{\varnothing}^B}=ts$, we can derive
\begin{align} \label{eq:t_fixed}
    t = r\sqrt{\frac{1-\delta^2+(1-r^2)\frac{\delta^2}{Q^2}}{1-r^2\delta^2}} \ .
\end{align}
and
\begin{align}\label{eq:s_fixed}
    s = \frac{\delta}{Q} \sqrt{\frac{1-r^2\delta^2}{1-\delta^2+(1-r^2)\frac{\delta^2}{Q^2}}} \ .
\end{align}

Therefore, Bob cannot encode information about $x$ in states with arbitrary purity or overlaps. 

\section{Semidefinite programs to bound the certifiable randomness} \label{app:sdps}

In this part of the supplemental material, we present the semidefinite programming tools used in the main text to provide with bounds on the certifiable randomness in both the measurement-device-independent and semi-device-independent cases.

\subsection{Measurement-device-independent min-entropy}

To certify the randomness in both measurement outcomes, we will maximise the probability that any malicious party (namely Eve) guesses their values. We consider partially characterised state preparations which are pure ($\rho_x=\ket{\psi_x}\bra{\psi_x}$) and with a bounded overlap ($|\braket{\psi_0 | \psi_1}| \geq \delta$). On the other hand, we consider that Bob's device has a bounded dimension $d$, and subjected to shared randomness with the eavesdropper. Namely, each prepare-and-measure round the measurement devices can take distinct strategies which we label by $\lambda$ according to a distribution $q_\lambda$ of shared randomness. Eve can then use this shared randomness to improve their guessing probability, which we can write as
\begin{align}
p_g :=& \sum_\lambda q_\lambda \underset{b}{\max}\left\{\Tr\left[\rho_{x^\ast} M_{b}^\lambda\right]\right\} \label{eq:pgB_sup} \ .
\end{align}
To cast the maximisation of $p_g$ as an semidefinite program (SDP), we use the following tricks. Firstly, we define the operators $\tilde{M}_{b}^\lambda := q_\lambda M_{b}^\lambda$. Secondly, we note that we can bound the relevant number of strategies by only considering those that yield the maximums in \eqref{eq:pgB_sup}. That is, strategy $\lambda$ will yield the maximal correlations for $b=\lambda$. Finally, given these changes, we bound $p_g$ as
\begin{align}
	\text{maximise} & \quad p_g = \sum_{\lambda} \Tr\left[\rho_{x^\ast} \tilde{M}^\lambda_\lambda\right]  \\
	\text{such that} & \quad \tilde{M}^\lambda_b \succeq 0 , \ \sum_{\lambda,b} \tilde{M}^\lambda_b = \mathds{1} , \nonumber \\
	& \quad \sum_b \tilde{M}^\lambda_b = \Tr\left[\sum_b \tilde{M}^\lambda_b\right] \frac{\mathds{1}}{d} \nonumber \\
	& \quad \sum_\lambda \Tr\left[\rho_x \tilde{M}_{b}^\lambda\right] = p(b|x)  \ . \nonumber
\end{align}
If instead of the full distribution $p(b|x)$ one aims to bound the maximum confidence $C$ and the rate of inconclusive events $Q$, the last constraint should be replaced by
\begin{align}
\frac{1}{n}\sum_{x,\lambda} \Tr\left[\rho_x \tilde{M}_{\varnothing}^\lambda\right] \geq Q \ , \quad \frac{1}{n}\sum_{x,\lambda}  \Tr\left[\rho_x \tilde{M}_{x}^\lambda\right] \geq C\left(1-Q\right) \ .
\end{align}
In what follows, we will stick to the case where the confidence and rate of inconclusive events are bounded. The Lagrangian reads
\begin{align}
\mathcal{L} =& \sum_{\lambda} \Tr\left[\rho_{x^\ast} \tilde{M}^\lambda_\lambda\right] + \Tr\left[\tilde{M}^\lambda_b W^\lambda_b\right] - \Tr\left[R\left(\sum_{\lambda,b} \tilde{M}^\lambda_b - \mathds{1}\right)\right] + \sum_\lambda\Tr\left[H^\lambda\left(\sum_b \tilde{M}^\lambda_b - \Tr\left[\sum_b \tilde{M}^\lambda_b\right] \frac{\mathds{1}}{d}\right)\right] \\
-& g\left(\frac{1}{n}\sum_{x,\lambda} \Tr\left[\rho_x \tilde{M}_{\varnothing}^\lambda\right] - Q\right) - h\left(\frac{1}{n}\sum_{x,\lambda} \Tr\left[\rho_x \tilde{M}_{x}^\lambda\right] - C\left(1-Q\right)\right)
\end{align}
Here we introduced the dual variables $d\times d$ matrices $W_b^\lambda$, $R$, $H^\lambda$ and the scalars $g$ and $h$ as Lagrangian multipliers. The dual formulation of the primal semidefinite program is obtained by minimizing over the dual variables the supremum $\mathcal{S}$ of the Lagrangian over the primal variables. Namely,
\begin{align}
    \mathcal{S} = \sup_{\tilde{M}^\lambda_b} \mathcal{L} = \sup_{\tilde{M}^\lambda_b} \left\{gQ+hC(1-Q)+\Tr\left[R\right] + \sum_{b,\lambda}\Tr\left[\tilde{M}^\lambda_b K_b^\lambda\right]\right\}
\end{align}
for
\begin{align}
    K_{b}^\lambda = \sum_{x} \rho_x \left(\delta_{b,\lambda}\delta_{x,x^\ast}-\frac{g}{n}\delta_{b,\varnothing}-\frac{h}{n}\delta_{b,x}\right) + H^{\lambda} - \frac{1}{d}\Tr\left[H^{\lambda}\right]\mathds{1} - R + W_b^\lambda
\end{align}
In order to avoid diverging solutions, we need to impose $K_{b}^\lambda=0$ or, equivalently, that  $K_{b}^\lambda - W_b^\lambda \preceq 0$, since $W_b^\lambda\succeq 0$. Therefore, the dual semidefinite program results form the minimisation of $\mathcal{S}$ under this condition. Namely, the dual reads
\begin{align} \label{eq:HminB_dual}
	\text{minimize} & \quad gQ+hC(1-Q)+\Tr\left[R\right]  \\
	\text{such that} & \quad \sum_{x} \rho_x \left(\delta_{b,\lambda}\delta_{x,x^\ast}-\frac{g}{n}\delta_{b,\varnothing}-\frac{h}{n}\delta_{b,x}\right) + H^{\lambda} - \frac{1}{d}\Tr\left[H^{\lambda}\right]\mathds{1} - R \preceq 0  \ . \nonumber
\end{align}
Here $g$ and $h$ are arbitrary scalar variables, and $R$ and $H^{\lambda}$ are $d\times d$ hermitian matrices.

\subsection{Semi-device-independent randomness: Shannon and min-entropies}

We define $G_{b,c}^{\lambda}:=K_b^\lambda N_c^\lambda (K_b^\lambda)^\dagger$ the POVM element representing Bob and Charlie's sequential measurement scheme. Their joint observed probabilities, together with their respective marginalisations, can be written as
\begin{align}
p(b,c|x) = \sum_\lambda q(\lambda) \Tr\left[\rho_x G_{b,c}^{\lambda}\right] \ , \quad p(b|x) = \sum_c p(b,c|x) \ , \quad p(c|x) = \sum_b p(b,c|x) \ .
\end{align}
On the other hand, if we bound the certifiable confidences with $C^{B}$ and $C^{C}$ and observed rates of inconcluisve events $p_{\varnothing}^{\rm B}\leq Q^{B}$ and $p_{\varnothing}^{\rm C}\leq Q^{C}$ in Bob and Charlie's devices respectively, one has
\begin{align}
    &\frac{1}{n}\sum_{x,c,\lambda} q(\lambda)\Tr\left[\rho_x G_{\varnothing,c}^\lambda\right] \geq Q^{B} \ , \quad \frac{1}{n}\sum_{x,c,\lambda}  q(\lambda)\Tr\left[\rho_x G_{x,c}^\lambda\right] \geq C^{B}\left(1-Q^{B}\right) \\
    &\frac{1}{n}\sum_{x,b,\lambda} q(\lambda)\Tr\left[\rho_x G_{b,\varnothing}^\lambda\right] \geq Q^{C} \ , \quad \frac{1}{n}\sum_{x,b,\lambda}  q(\lambda)\Tr\left[\rho_x G_{b,x}^\lambda\right] \geq C^{C}\left(1-Q^{C}\right)
\end{align}
In the following we present a method to bound the certifiable Shannon and min-entropies under bounded confidences and rates of inconclusive events in both devices individually.

\subsubsection{Min-entropy}

The probability that Eve guesses both measurement outcomes can be written as
\begin{align}
p_g^{BC} = \sum_{\lambda} q(\lambda) \ \underset{b,c}{\max}\left\{\Tr\left[\rho_{x^\ast} G^{\lambda}_{b,c}\right]\right\} \ . \label{eq:pgBCapp}
\end{align}
We aim to maximize $p_g^{BC}$ for any set of implementable operations $G^{\lambda}_{b,c}$ and distributions $q(\lambda)$ such that the observed confidences $C^{B}$ and $C^{C}$ and bounds on the rates of inconclusive events $p_{\varnothing}^{\rm B}\geq Q^{B}$ and $p_{\varnothing}^{\rm C}\geq Q^{C}$ are satisfied in Bob and Charlie's devices, respectively. In order to cast this optimization as a semidefinite program, we need to linearize the problem. First, we define the new positive-semidefinite operators $\tilde{G}^{\lambda}_{b,c}:=q(\lambda)G^{\lambda}_{b,c}$. Normalisation of the original $G^{\lambda}_{b,c}$ implies
\begin{align}
\sum_{b,c} \tilde{G}^{\lambda}_{b,c} = \Tr\left[\sum_{b,c} \tilde{G}^{\lambda}_{b,c}\right] \frac{\mathds{1}}{d} \ .
\end{align}
Second, we reduce the number of strategies labelled by the shared randomness $\lambda$ to those that are relevant in our optimisation. Namely, we only consider those strategies that yield maximal correlations for each outcome $b$ and $c$. We define the tuple $\lambda=(\lambda_1,\lambda_2)$ and call $\lambda_1=b$ and $\lambda_2=c$ the strategies that give the correlations that satisfy the maximum in \eqref{eq:pgBCapp}. With all that, we can write the optimisation problem in a semidefinite program form,
\begin{align}
	\text{maximise} & \quad p_g^{BC} = \sum_{\lambda_1 \lambda_2} \Tr\left[\rho_{x^\ast} \tilde{G}^{\lambda_1 \lambda_2}_{\lambda_1 \lambda_2}\right]  \\
	\text{such that} & \quad \tilde{G}^{\lambda_1 \lambda_2}_{b,c} \succeq 0 , \ \sum_{\lambda_b \lambda_c,b,c} \tilde{G}^{\lambda_1 \lambda_2}_{b,c} = \mathds{1} , \nonumber \\
	& \quad \sum_{b,c} \tilde{G}^{\lambda_1 \lambda_2}_{b,c} = \Tr\left[\sum_{b,c} \tilde{G}^{\lambda_1 \lambda_2}_{b,c}\right] \frac{\mathds{1}}{d} \nonumber \\
    & \quad \frac{1}{n}\sum_{x,c,\lambda_1,\lambda_2} \Tr\left[\rho_x \tilde{G}_{\varnothing,c}^{\lambda_1,\lambda_2}\right] \geq Q^{B} \ , \quad \frac{1}{n}\sum_{x,c,\lambda_1,\lambda_2}  \Tr\left[\rho_x \tilde{G}_{x,c}^{\lambda_1,\lambda_2}\right] \geq C^{B}\left(1-Q^{B}\right) \nonumber \\
    & \quad \frac{1}{n}\sum_{x,b,\lambda_1,\lambda_2} \Tr\left[\rho_x \tilde{G}_{b,\varnothing}^{\lambda_1,\lambda_2}\right] \geq Q^{C} \ , \quad \frac{1}{n}\sum_{x,b,\lambda_1,\lambda_2}  \Tr\left[\rho_x \tilde{G}_{b,x}^{\lambda_1,\lambda_2}\right] \geq C^{C}\left(1-Q^{C}\right)   \ . \nonumber
\end{align}
We proceed providing the dual form of the semidefinite program. The Lagrangian is
\begin{align}
\mathcal{L} =& \sum_{\lambda_1 \lambda_2} \Tr\left[\rho_{x^\ast} \tilde{G}^{\lambda_1 \lambda_2}_{\lambda_1 \lambda_2}\right] + \Tr\left[\tilde{G}^{\lambda_1 \lambda_2}_{b,c}W^{\lambda_1 \lambda_2}_{b,c}\right] - \Tr\left[R\left(\tilde{G}^{\lambda_1 \lambda_2}_{b,c} - \mathds{1}\right)\right] + \sum_{\lambda_1 \lambda_2}\Tr\left[ F^{\lambda_1 \lambda_2} \left(\sum_{b,c} \tilde{G}^{\lambda_1 \lambda_2}_{b,c} - \Tr\left[\sum_{b,c} \tilde{G}^{\lambda_1 \lambda_2}_{b,c}\right]\frac{\mathds{1}}{d}\right) \right] \\
-& g^{\rm B}\left(\frac{1}{n}\sum_{x,c,\lambda_1,\lambda_2} \Tr\left[\rho_x \tilde{G}_{\varnothing,c}^{\lambda_1,\lambda_2}\right] - Q^{B}\right) - h^{\rm B}\left(\frac{1}{n}\sum_{x,c,\lambda_1,\lambda_2}  \Tr\left[\rho_x \tilde{G}_{x,c}^{\lambda_1,\lambda_2}\right] - C^{B}\left(1-Q^{B}\right)\right) \nonumber \\
-& g^{\rm C}\left(\frac{1}{n}\sum_{x,b,\lambda_1,\lambda_2} \Tr\left[\rho_x \tilde{G}_{b,\varnothing}^{\lambda_1,\lambda_2}\right] - Q^{C}\right) - h^{\rm C}\left(\frac{1}{n}\sum_{x,b,\lambda_1,\lambda_2}  \Tr\left[\rho_x \tilde{G}_{b,x}^{\lambda_1,\lambda_2}\right] - C^{C}\left(1-Q^{C}\right)\right)  \nonumber \ , 
\end{align}
where we introduced the dual positive-semidefinite variables $W^{\lambda_1 \lambda_2}_{b,c}$ together with the Hermitian $d\times d$ matrices $R$, $F^{\lambda_1 \lambda_2}$ and the scalar variables $g^{\rm B}$, $g^{\rm C}$, $h^{\rm B}$ and $h^{\rm C}$. The dual here is constructed by minimizing over the dual variables the supremum $\mathcal{S}$ of the Lagrangian over the primal variables, i.e.
\begin{align}
    \mathcal{S} = \sup_{\tilde{G}^{\lambda_1 \lambda_2}_{b,c}} \mathcal{L} = \sup_{\tilde{G}^{\lambda_1 \lambda_2}_{b,c}} \left\{g^{\rm B} p_{\varnothing}^{\rm B}+h^{\rm B} C^{\rm B}(1-p_{\varnothing}^{\rm B})+g^{\rm C} p_{\varnothing}^{\rm C}+h^{\rm C} C^{\rm C}(1-p_{\varnothing}^{\rm C})+\Tr\left[R\right] + \sum_{b,c,\lambda_1,\lambda_2}\Tr\left[\tilde{G}^{\lambda_1 \lambda_2}_{b,c} K_{b,c}^{\lambda_1,\lambda_2}\right]\right\} \ ,
\end{align}
for
\begin{align}
    K_{b,c}^{\lambda_1,\lambda_2} = \sum_{x} \rho_x \left(\delta_{b,\lambda_1}\delta_{c,\lambda_2}\delta_{x,x^\ast}-\frac{g^{\rm B}}{n}\delta_{b,\varnothing}-\frac{h^{\rm B}}{n}\delta_{b,x}-\frac{g^{\rm C}}{n}\delta_{c,\varnothing}-\frac{h^{\rm C}}{n}\delta_{c,x}\right) + F^{\lambda_1 \lambda_2} - \frac{1}{2}\Tr\left[F^{\lambda_1 \lambda_2}\right]\mathds{1} - R + W^{\lambda_1 \lambda_2}_{b,c} \ .
\end{align}
To avoid diverging solutions, we need to impose $K_{b,c}^{\lambda_1,\lambda_2}=0$, or equivalently $K_{b,c}^{\lambda_1,\lambda_2}-W^{\lambda_1 \lambda_2}_{b,c} \preceq 0$ given that $W^{\lambda_1 \lambda_2}_{b,c}\succeq 0$. Therefore, the dual semidefinite program is recovered from the minimization of the suppremum $\mathcal{S}$ under this condition. That is,
\begin{align}
	p_g^{\rm BC}\leq \text{minimize} & \quad g^{\rm B} p_{\varnothing}^{\rm B}+h^{\rm B} C^{\rm B}(1-p_{\varnothing}^{\rm B})+g^{\rm C} p_{\varnothing}^{\rm C}+h^{\rm C} C^{\rm C}(1-p_{\varnothing}^{\rm C})+\Tr\left[R\right]  \\
	\text{such that} & \quad \sum_{x} \rho_x \left(\delta_{b,\lambda_1}\delta_{c,\lambda_2}\delta_{x,x^\ast}-\frac{g^{\rm B}}{n}\delta_{b,\varnothing}-\frac{h^{\rm B}}{n}\delta_{b,x}-\frac{g^{\rm C}}{n}\delta_{c,\varnothing}-\frac{h^{\rm C}}{n}\delta_{c,x}\right) + F^{\lambda_1 \lambda_2} - \frac{1}{d}\Tr\left[F^{\lambda_1 \lambda_2}\right]\mathds{1} - R \preceq 0
\end{align}
Similarly, the dual semidefinite program to bound the individual certifiable randomness only in Charlie's device is
\begin{align}
	p_g^{\rm C}\leq \text{minimise} & \quad g^{\rm B} p_{\varnothing}^{\rm B}+h^{\rm B} C^{\rm B}(1-p_{\varnothing}^{\rm B})+g^{\rm C} p_{\varnothing}^{\rm C}+h^{\rm C} C^{\rm C}(1-p_{\varnothing}^{\rm C})+\Tr\left[R\right]  \\
	\text{such that} & \quad \sum_{x} \rho_x \left(\delta_{c,\lambda}\delta_{x,x^\ast}-\frac{g^{\rm B}}{n}\delta_{b,\varnothing}-\frac{h^{\rm B}}{n}\delta_{b,x}-\frac{g^{\rm C}}{n}\delta_{c,\varnothing}-\frac{h^{\rm C}}{n}\delta_{c,x}\right) + F^{\lambda} - \frac{1}{d}\Tr\left[F^{\lambda}\right]\mathds{1} - R \preceq 0 \ .
\end{align}

\subsubsection{Shannon entropy}

Following Refs.~\cite{brown2021,carceller2025improving,carceller2025photon}, we can write the Shannon entropy as
\begin{equation}\label{eq:sb}
    H(BC|E) \geq c_m + \sum_{i=1}^{m-1}\tau_i \sum_{jk} \ \underset{z_{ijk}}{\inf} \left\{ \Tr\left[\rho_{x^{*}} G_{j,k}\right]\left(2z_{ijk}+ (1-t_i)z_{ijk}^2\right)+ t_i z_{ijk}^2 \right\}\,.
\end{equation}
where $\tau_{i}:=\frac{w_{i}}{t_{i}\ln{2}}$ and $c_m:=\sum_{i=0}^{m-1}\tau_i$ are defined in term of the nodes $t_i$ and weights $w_i$of the Gauss-Radau quadratures (see \cite{gautschi2000,brown2021}), and $z_{ijk}$ are real scalar variables. 

We define the variables $K^{i,j,k}_{b,c}(n):=z_{ijk}^nG_{b,c}$, for $n$ being the power of the scalar $z_{ijk}$. The primal semidefinite program reads
\begin{align}
    \underset{\left\{G_{b,c},K^{i,j,k}_{b,c}(n)\right\}}{\text{minimize}} & \quad c_m + \sum_{i=0}^{m-1} \sum_{jk} \tau_i \Tr\left[\rho_{x^{*}}\left(2K^{i,j,k}_{j,k}(1)+(1-t_i)K^{i,j,k}_{j,k}(2) + t_i \sum_{b,c}K^{i,j,k}_{b,c}(2)\right)\right] \\    
    \text{subject to} \quad & \quad \begin{pmatrix}
    G_{b,c} & K^{i,j,k}_{b,c}(1) \\
    K^{i,j,k}_{b,c}(1) & K^{i,j,k}_{b,c}(2)
\end{pmatrix}     \succeq 0, \quad \sum_{b,c}G_{b,c} = \mathds{1}, \quad \sum_{b,c}K^{i,j,k}_{b,c}(n) = \frac{1}{d}\Tr\left[\sum_{b}K^{i,j,k}_{b,c}(n)\right]\mathds{1} \nonumber \\
    & \quad \frac{1}{n}\sum_{x,c} \Tr\left[\rho_x G_{\varnothing,c}\right] \geq Q^{B} \ , \quad \frac{1}{n}\sum_{x,c}  \Tr\left[\rho_x G_{x,c}\right] \geq C^{B}\left(1-Q^{B}\right) \nonumber \\
    & \quad \frac{1}{n}\sum_{x,b} \Tr\left[\rho_x G_{b,\varnothing}\right] \geq Q^{C} \ , \quad \frac{1}{n}\sum_{x,b}  \Tr\left[\rho_x G_{b,x}\right] \geq C^{C}\left(1-Q^{C}\right)   \ . \nonumber
\end{align}

We continue deriving the dual form of the semidefinite program above. The Lagrangian reads
\begin{align}
\mathcal{L} =& c_m + \sum_{i=0}^{m-1} \sum_{jk} \tau_i \Tr\left[\rho_{x^{*}}\left(2K^{i,j,k}_{j,k}(1)+(1-t_i)K^{i,j,k}_{j,k}(2) + t_i \sum_{b,c}K^{i,j,k}_{b,c}(2)\right)\right] \\
+& \Tr\left[G_{b,c}D^{ijk}_{b,c} + K^{i,j,k}_{j,k}(1)\left(F^{i,j,k}_{b,c}+F^{i,j,k\dagger}_{b,c}\right) + K^{i,j,k}_{j,k}(2)L^{ijk}_{b,c}\right] - \Tr\left[R\left(\sum_{b,c}G_{b,c} - \mathds{1}\right)\right]  \nonumber \\
+& \sum_{i,j,k,n}\Tr\left[Q^{i,j,k}_n\left(\sum_{b,c}K^{i,j,k}_{b,c}(n) - \frac{1}{d}\Tr\left[\sum_{b}K^{i,j,k}_{b,c}(n)\right]\mathds{1}\right)\right] \nonumber \\
-& g^{\rm B}\left(\frac{1}{n}\sum_{x,c} \Tr\left[\rho_x G_{\varnothing,c}\right] - Q^{B}\right) - h^{\rm B} \left(\frac{1}{n}\sum_{x,c}  \Tr\left[\rho_x G_{x,c}\right] - C^{B}\left(1-Q^{B}\right)\right) \nonumber \\
- & g^{\rm C}\left( \frac{1}{n}\sum_{x,b} \Tr\left[\rho_x G_{b,\varnothing}\right] - Q^{C}\right) - h^{\rm C}\left(\frac{1}{n}\sum_{x,b}  \Tr\left[\rho_x G_{b,x}\right] - C^{C}\left(1-Q^{C}\right)\right)   \ , \nonumber
\end{align}
where we introduced the $d\times d$ dual variables $D_{bc}^{ijk}$, $F_{bc}^{ijk}$, $L_{bc}^{ijk}$, $R$, $Q_1^{ijk}$ and $Q_2^{ijk}$ and scalars $g^{\rm B}$, $h^{\rm B}$, $g^{\rm C}$ and $h^{\rm C}$ as Lagrangian multipliers. To retrieve the dual semidefinite program, we need to maximize over the dual variables the infimum $\mathcal{I}$ of the Lagrangian over the primal variables. That is,
\begin{align}
    \mathcal{I} = \inf_{\left\{G_{b,c},K^{i,j,k}_{b,c}(n)\right\}} \mathcal{L} = \inf_{\left\{G_{b,c},K^{i,j,k}_{b,c}(n)\right\}} \left\{ c_m - g p_{\varnothing}^{\rm B}-fC^{\rm B}(1-p_{\varnothing}^{\rm B})-\Tr\left[R\right] + \sum_{b,c}\Tr\left[G_{b,c}K_{b,c}\right] + \sum_{b,c,i,j,k,n}\Tr\left[K^{i,j,k}_{b,c}(n)S^{i,j,k}_{b,c}(n)\right] \right\} \ ,
\end{align}
for
\begin{align}
    K_{b,c} &= \sum_{x}\rho_x\left(\frac{g^{\rm B}}{n}\delta_{b,\varnothing}+\frac{h^{\rm B}}{n}\delta_{b,x} + \frac{g^{\rm C}}{n}\delta_{c,\varnothing}+\frac{h^{\rm C}}{n}\delta_{c,x}\right) + R - \sum_{ijk} D_{bc}^{ijk} \\
    S^{i,j,k}_{b,c}(1) &= 2\tau_i\rho_{x^\ast}\delta_{j,b}\delta_{k,c} + Q_1^{ijk} - \frac{1}{d}\Tr\left[Q_1^{ijk}\right]\mathds{1} - F_{bc}^{ijk} - F_{bc}^{ijk\dagger} \\
    S^{i,j,k}_{b,c}(2) & = \tau_i\rho_{x^\ast}\!\left[\left(1\!-\!t_i\right)\delta_{j,b}\delta_{k,c}\!+\!t_i\right]\! +\! Q_2^{ijk}\! -\! \frac{1}{d}\Tr\left[Q_2^{ijk}\right]\mathds{1} - L_{bc}^{ijk} \ .
\end{align}
To avoid diverging solutions, the dual semidefinite program is retrieved with the maximisation of the infimum $\mathcal{I}$ under the conditions that $K_{b,c}=0$ and $S^{i,j,k}_{b,c}(n)=0$. This results into the dual formulation, which reads
\begin{align}
    H(BC|E)\geq\text{maximize} & \quad c_m - g^{\rm B} p_{\varnothing}^{\rm B}-h^{\rm B}C^{\rm B}(1-p_{\varnothing}^{\rm B}) - g^{\rm C} p_{\varnothing}^{\rm C}-h^{\rm C}C^{\rm C}(1-p_{\varnothing}^{\rm C}) - \Tr\left[R\right] \\  
    \text{subject to} & \quad \begin{pmatrix}
        D_{bc}^{ijk} & F_{bc}^{ijk} \\
        F_{bc}^{ijk\dagger} & L_{bc}^{ijk}
    \end{pmatrix} \succeq 0  \nonumber \\
    & \quad \sum_{ijk} D_{bc}^{ijk} = \sum_{x}\rho_x\left(\frac{g^{\rm B}}{n}\delta_{b,\varnothing}+\frac{h^{\rm B}}{n}\delta_{b,x} + \frac{g^{\rm C}}{n}\delta_{c,\varnothing}+\frac{h^{\rm C}}{n}\delta_{c,x}\right) + R \nonumber   \\
& \quad F_{bc}^{ijk} + F_{bc}^{ijk\dagger} = 2\tau_i\rho_{x^\ast}\delta_{j,b}\delta_{k,c} + Q_1^{ijk} - \frac{1}{d}\Tr\left[Q_1^{ijk}\right]\mathds{1} \nonumber \\
& \quad L_{bc}^{ijk}\!=\!\tau_i\rho_{x^\ast}\!\left[\left(1\!-\!t_i\right)\delta_{j,b}\delta_{k,c}\!+\!t_i\right]\! +\! Q_2^{ijk}\! -\! \frac{1}{d}\Tr\left[Q_2^{ijk}\right]\mathds{1}   \nonumber ,
\end{align}
One can similalry follow all the same process to obtain the dual semidefinite programs to obtain bounds on the Shannon entropy individually in Bob and Charlie's devices. Concretely, the Shannon entropy in Bob's device is bounded through the following semidefinite program,
\begin{align}
    H(B|E)\geq\text{maximize} & \quad c_m - g p_{\varnothing}^{\rm B}-hC^{\rm B}(1-p_{\varnothing}^{\rm B})-\Tr\left[R\right] \\    
    \text{subject to} & \quad \begin{pmatrix}
        D_{b}^{ij} & F_{b}^{ij} \\
        F_{b}^{ij\dagger} & L_{b}^{ij}
    \end{pmatrix} \succeq 0  \nonumber \\
    & \quad \sum_{ij} D_{b}^{ij} = \sum_{x}\rho_x \left(\frac{g}{n}\delta_{b,\varnothing}+\frac{h}{n}\delta_{b,x}\right) 
     + R \nonumber   \\
& \quad F_{b}^{ij} + F_{b}^{ij\dagger} = 2\tau_i\rho_{x^\ast}\delta_{j,b} + Q_1^{ij} - \frac{1}{d}\Tr\left[Q_1^{ij}\right]\mathds{1} \nonumber \\
& \quad L_{b}^{ij}\!=\!\tau_i\rho_{x^\ast}\!\left[\left(1\!-\!t_i\right)\delta_{j,b}\!+\!t_i\right]\! +\! Q_2^{ij}\! -\! \frac{1}{d}\Tr\left[Q_2^{ij}\right]\mathds{1}   \nonumber \ .
\end{align}
Charlie's Shannon entropy is bounded through the semidefinite program,
\begin{align}
    H(C|E)\geq\text{maximize} & \quad c_m - g^{\rm B} p_{\varnothing}^{\rm B}-h^{\rm B}C^{\rm B}(1-p_{\varnothing}^{\rm B}) - g^{\rm C} p_{\varnothing}^{\rm C}-h^{\rm C}C^{\rm C}(1-p_{\varnothing}^{\rm C}) - \Tr\left[R\right] \\    
    \text{subject to} & \quad \begin{pmatrix}
        D_{bc}^{ij} & F_{bc}^{ij} \\
        F_{bc}^{ij\dagger} & L_{bc}^{ij}
    \end{pmatrix} \succeq 0  \nonumber \\
    & \quad \sum_{ij} D_{bc}^{ij} = \sum_{x}\rho_x\left(\frac{g^{\rm B}}{n}\delta_{b,\varnothing}+\frac{h^{\rm B}}{n}\delta_{b,x} + \frac{g^{\rm C}}{n}\delta_{c,\varnothing}+\frac{h^{\rm C}}{n}\delta_{c,x}\right) + R \nonumber   \\
& \quad F_{bc}^{ij} + F_{bc}^{ij\dagger} = 2\tau_i\rho_{x^\ast}\delta_{j,c} + Q_1^{ij} - \frac{1}{d}\Tr\left[Q_1^{ij}\right]\mathds{1} \nonumber \\
& \quad L_{bc}^{ij}\!=\!\tau_i\rho_{x^\ast}\!\left[\left(1\!-\!t_i\right)\delta_{j,c}\!+\!t_i\right]\! +\! Q_2^{ij}\! -\! \frac{1}{d}\Tr\left[Q_2^{ij}\right]\mathds{1}   \nonumber
\end{align}

\section{Limiting inconclusive rate values for randomness generation} \label{app:inc_lims}

In this section of the supplemental material, we derive the critical inconclusive rates at which randomness cannot be certified. 

To certify the randomness of the measurement outcomes, we need to make sure that no predefined measurements can deterministically reproduce the correlations observed in the prepare-and-measure scenario. Namely, if a set of strategies $\lambda$ distributed according to $\lambda$ can be found such that
\begin{align}
    p(b|x) = \sum_{\lambda} q(\lambda) p_\lambda(b|x) \ ,
\end{align}
for $p_\lambda(b|x)$ being the specific correlations reproduced by the deterministic strategy $\lambda$, then randomness cannot be certified.

We begin looking at the case of unambiguous state discrimination of two pure states. Consider two state preparations of the form $\ket{\psi_x}$ for $x=0,1$ with an overlap $\braket{\psi_0|\psi_1}=\delta$. To satisfy the no-error condition, any deterministic measurement strategy must by aligned with the state preparations. Since we consider only two state preparations, we are bound to have the following two strategies
\begin{align}
   & \text{for:} \ \lambda=0 && \text{for:} \ \lambda=1 \\
   & M_0^0 = 0 && M_0^1 = (2-c)\ket{\psi_1^\perp}\bra{\psi_1^\perp} \\
   & M_1^0 = (2-c)\ket{\psi_0^\perp}\bra{\psi_0^\perp} && M_1^1 = 0 \\
   & M_{\varnothing}^0 = (2-c)\ket{\psi_0}\bra{\psi_0} + (c-1)\mathds{1} && M_{\varnothing}^1 = (2-c)\ket{\psi_1}\bra{\psi_1} + (c-1)\mathds{1} \ .
\end{align}
The parameter $c$ is limited by $1\leq c \leq 2$ to ensure positivity of the POVM elements. The rate of inconclusive events is given by
\begin{align}
    p_{\varnothing} = \sum_\lambda q_\lambda p_{\varnothing}^\lambda = \sum_\lambda q_\lambda \frac{1}{2}\left(\bra{\psi_0}M_{\varnothing}^\lambda\ket{\psi_0}+\bra{\psi_1}M_{\varnothing}^\lambda\ket{\psi_1}\right) = \delta^2 + \frac{1-\delta^2}{2}c \ ,
\end{align}
which is limited by $\frac{1+\delta^2}{2}\leq p_{\varnothing}\leq 1$.

Let's now generalize to maximum confidence measurements. Here, the no-error condition from unambiguous state discrimination is translated to a no-error condition for complementary state preparations $\ket{\varphi_x}$. Any deterministic measurement strategy, therefore, must be aligned with those. Take the two deterministic measurement strategies
\begin{align}
   & \text{for:} \ \lambda=0 && \text{for:} \ \lambda=1 \\
   & M_0^0 = 0 && M_0^1 = (2-c)\ket{\varphi_1^\perp}\bra{\varphi_1^\perp} \\
   & M_1^0 = (2-c)\ket{\varphi_0^\perp}\bra{\varphi_0^\perp} && M_1^1 = 0 \\
   & M_{\varnothing}^0 = (2-c)\ket{\varphi_0}\bra{\varphi_0} + (c-1)\mathds{1} && M_{\varnothing}^1 = (2-c)\ket{\varphi_1}\bra{\varphi_1} + (c-1)\mathds{1} \ .
\end{align}
The rate of inconclusive events is given by
\begin{align}
    p_{\varnothing} = \sum_\lambda q_\lambda p_{\varnothing}^\lambda = \sum_\lambda q_\lambda \frac{1}{2}\left(\bra{\psi_0}M_{\varnothing}^\lambda\ket{\psi_0}+\bra{\psi_1}M_{\varnothing}^\lambda\ket{\psi_1}\right) = r^2\delta^2 + \frac{1-r^2\delta^2}{2}c \ ,
\end{align}
which is limited by $\frac{1+r^2\delta^2}{2}\leq p_{\varnothing}\leq 1$ since $1\leq c \leq 2$.

Interestingly, this result is related to simulating measurements with classical strategies. Namely, if $p_{\varnothing}\leq\frac{1+r^2\delta^2}{2}$, the observable correlations with maximum confidence measurements cannot be simulated with deterministic measurement strategies.

\end{widetext}

\end{document}